\documentclass[oneside,12pt,a4paper]{article}
\usepackage [latin1] {inputenc}
\usepackage{amsmath}
\usepackage{amssymb}
\usepackage{amsfonts}
\usepackage{exscale}
\usepackage{color}
\usepackage[authoryear, round]{natbib}
\usepackage{theorem}
\usepackage{graphicx}
\usepackage{setspace}
\usepackage{subcaption}
\usepackage{enumerate}

\newenvironment{myproof}[2] {\paragraph{Proof of {#1} {#2}:}}{\hfill$\square$}

\newtheorem{lem}{Lemma}
\newtheorem{thm}{Theorem}
\newtheorem{cor}{Corollary}
\newtheorem{prop}{Proposition}
\newtheorem{assumption}{Assumption}

\newenvironment{keywords}%
   {\begin{trivlist}\item[]{\bfseries Keywords:}\ }
   {\end{trivlist}}

\newenvironment{JEL}%
   {\begin{trivlist}\item[]{\bfseries JEL Classification:}\ }
   {\end{trivlist}}

\newcommand{\cA}{\mathcal{A}}
\newcommand{\cS}{\mathcal{S}}
\newcommand{\Diag}{\operatorname{Diag}}
\newcommand{\R}{\operatorname{\mathbb{R}}}

\begin{document}

\author{Anne G. Balter~~~~~~~Nikolaus Schweizer~~~~~~~Juan C. Vera\thanks{Department of Econometrics and OR, Tilburg School of Economics and Management, Tilburg University, Tilburg, The Netherlands. \{a.g.balter, n.f.f.schweizer, j.c.veralizcano\}@uvt.nl }\\}

\vspace{2.cm}

\title{\textbf{Contingent Capital with Stock Price Triggers in Interbank Networks}\thanks{We thank Sven Balder, Diego Ronchetti and Harald Uhlig and participants at the 2020 Econometric Society World Congress for helpful comments.}}
\date{November 2020}
\maketitle

\begin{abstract}

This paper studies existence and uniqueness of equilibrium prices in a model of the banking sector in which banks trade contingent convertible bonds with stock price triggers among each other. This type of financial product was proposed as an instrument for stabilizing the global banking system after the financial crisis. Yet it was recognized early on that these products may create circularity problems in the definition of stock prices -- even in the absence of trade. We find that if conversion thresholds are such that bond holders are indifferent about marginal conversions, there exists a unique equilibrium irrespective of the network structure. When thresholds are lower, existence of equilibrium breaks down while higher thresholds may lead to multiplicity of equilibria. Moreover, there are complex network effects. One bank's conversion may trigger further conversions -- or prevent them, depending on the constellations of asset values and conversion triggers.\bigskip

\end{abstract}

\doublespacing
\newpage
\section{Introduction}

Over the past decade, there has been considerable interest in contingent convertible debt as an instrument for making the global banking system more resilient towards crisis. The idea behind such contingent convertible bonds (CoCos) is simple. Banks issue bonds, i.e., borrow money, that is equipped with a conversion feature. Whenever the bank is in distress and there is sufficient danger that it cannot pay back this debt, the conversion occurs. After the conversion, the bank's former creditors own part of the bank -- but they do not receive their money back. The purpose of CoCos is thus to provide liquidity to banks in a way that does not endanger the stability of the banking system and reduces systemic risk. Yet once a new product type is introduced and traded between financial institutions, crossholdings in this product become a source of additional interconnectedness in the financial market. In this paper, we study how this interconnectedness may affect
the valuation of banks and create complex interdependencies between the potential conversions of different CoCos. Once banks hold each others' CoCos, conversion events of one bank may trigger or prevent conversions and defaults of other banks.

In the design of CoCos, the definition of the conversion event is critical. If conversion happens only when the borrowing bank is close to bankrupt, the conversion event can be expected to affect the banking system almost like a default: Instead of getting its money back, the lending institution becomes partial owner of an almost bankrupt company. In contrast, if conversion happens to a borrowing bank which is in excellent shape, the lenders may actually be better off being compensated in stocks than receiving their money. In this case, the conversion event corresponds to a wealth transfer from the existing owners to the lenders which goes together with partially losing control of the company. The threshold level for conversions should thus neither be chosen too high nor too low.

In addition to choosing the level of the conversion event, the designer of a CoCo bond needs to specify which performance indicators of the borrowing institution are used as trigger quantities. For instance, conversion could be based on accounting figures found in the bank's balance sheet. While this approach is relatively transparent, it may be too slow and prone to creative accounting. Alternatively, the conversion decision could be put into the hands of a regulating authority. In this case, the timing of conversions becomes more flexible but also rather intransparent. Moreover, a conversion that is triggered by a regulator will be viewed by the market as a sign of distress and may lead to further disruptions.

We focus on a third possibility: stock prices as conversion triggers. Conversion thus occurs as soon as the stock price hits a prespecified lower bound. Compared to accounting-based or regulatory triggers, stock price triggers are more transparent and react much more quickly. Their potential drawback is that they may introduce a circularity in the definition of the stock price. The stock price reflects what it is worth to own part of the bank. Yet what that is worth depends on whether conversion happens or not -- which depends itself on the stock price. This observation has been the starting point of \cite{SW} who argue that neither existence nor uniqueness of equilibrium stock prices is guaranteed after the introduction of contingent convertible debt with a stock price trigger. \cite{flannery2014contingent} argues that this potential danger has been the main reason why stock price triggers have not been implemented in practice.\footnote{In practice, most CoCos have a combination of accounting-based and regulatory trigger mechanisms, see \cite{Avdjiev2}. Even in the absence of stock price triggers, CoCos are still viewed with some suspicion by many, see, e.g., the press reports about a recent incident involving CoCos issued by Santander such as ``When Investing Is About the CEO's Goodwill'' in the Wall Street Journal of 17 February  2019.} This is despite the fact that later research \citep{GN, PT1, PT2} has identified weak and transparent sufficient conditions that guarantee existence and uniqueness of equilibrium stock prices. However,  this literature has studied single CoCos in isolation, abstracting from the fact that whenever a product is sold there is someone who buys it.
 
We ask what happens if banks do not only issue CoCos with stock price triggers but also trade them among each other.  In particular, we study how previous results on existence and uniqueness of equilibrium prices generalize to the multi-bank case. We aim at understanding whether network effects can amplify the circularity problems that threaten existence of equilibrium even in the single-bank setting.  Given that CoCos were proposed with the goal of stabilizing the global banking system, this question is relevant both from an academic and from a regulatory perspective.\footnote{Empirically, there is conflicting evidence how prevalent crossholdings of CoCos are. Comparing the different sources discussed in \cite{Avdjiev1,Avdjiev2} and \cite{BW}, it seems fair to conclude that, in the large European market, the fraction of CoCos owned by other banks is between 5\% and 50\% depending on the precise context and data that are considered. These are fairly high numbers given that the current Basel regulation tends to discourage such crossholdings, possibly out of systemic risk concerns \citep{Avdjiev1}.}

Our theoretical framework generalizes the static baseline model of \cite{SW} and \cite{GN} to multiple, interconnected banks. Let us summarize our main results on existence and uniqueness of equilibrium. We say that a bank has set a \textit{fair} conversion threshold if both a bank's creditors and its stock-holders are indifferent between conversion and non-conversion when the stock is exactly at the conversion threshold. With a fair conversion threshold, marginal conversions do not lead to wealth transfers. If all CoCos in the market have fair conversion thresholds, a unique vector of equilibrium stock prices exists. If all conversion thresholds are super-fair in the sense of being at or above the fair threshold, existence of a possibly non-unique equilibrium is guaranteed. Yet as soon as one bank sells CoCos with lower, sub-fair conversion thresholds, existence of equilibrium stock prices is in danger not only for this bank but potentially for the entire banking system. In the single bank case, these results simplify, of course, to those of \cite{GN} where a fair threshold implies existence of a unique equilibrium, a sub-fair threshold implies non-existence and a super-fair threshold implies non-uniqueness. 

We thus find that existence and uniqueness of equilibria does not depend on the network structure. It only depends on the conversion thresholds that banks have set. A regulator who is solely interested in whether stock prices are well-defined could evaluate the situation of one bank after the other in isolation. Yet, of course, this does not imply that crossholdings in contingent capital cannot create interdependencies between conversions of different banks. Indeed, under fair conversion thresholds, there may be domino effects where the conversion of one CoCo bond weakens the position of the holders of that bond. Consequently, the holders' own CoCos may have to convert in equilibrium, potentially triggering further conversions. In the super-fair case, the situation is similar but more complex. Depending on the exact constellation of asset values and CoCo holdings, the conversion of one bank's CoCo may either help or harm other banks, thus triggering or preventing additional conversions.\footnote{This dichotomy we find in our model is in contrast to the earlier literature which focused on negative externalities that banks exert on each other through their conversions \citep{CW, BW}. To understand why conversions can prevent other conversions, recall that, in the super-fair case, marginal cases of conversions are beneficial to the holders of a CoCo compared to receiving the original debt.} There may thus be situations where exactly one bank has to convert in equilibrium to save all others -- but it is not uniquely determined which bank that is.

One main conclusion from these findings is that the structure of the crossholdings network matters when it comes to CoCos. Thus, a regulator who considers loosening the capital requirements for interbank crossholdings should at the same time gather sufficient information about CoCo ownership. As noted in \citet{Avdjiev2} and \citet{BW}, information about CoCo investment is not systematically collected by regulators so far. Without this information, an interconnected CoCo market will be not be transparent and potential spillovers will be hard to predict.

Our ultimate interest is in the (potentially ill-defined) mapping that computes equilibrium stock prices from asset values. A main technical insight is that the mapping
in the opposite direction, from stock prices to the underlying asset values, is much easier to understand and analyze. From realized stock prices, we can easily read off which banks are healthy, converting or bankrupt. Given this information, the relation between asset values and stock prices is explicit and linear. Much of our analysis is thus based on deriving properties of this mapping from stock prices to asset values which we call $\Phi$. When $\Phi$ is surjective, every vector of asset values has an associated equilibrium stock price. When $\Phi$ is injective, this stock price is unique. When $\Phi$ fails to be surjective, there exist vectors of asset values without a corresponding vector of stock prices.

In the fair case, the mapping $\Phi$ is continuous and the problem of identifying the unique equilibrium is reminiscent of the problem of finding equilibria in financial networks under interbank lending and default as discussed, e.g., in  \cite{acemoglu2015systemic}, \cite{glasserman2016contagion} and the references therein. The main difference is that our model has three states as each bank can be healthy, bankrupt or converting. Nevertheless, existence and uniqueness of equilibrium follow from classical results for continuous functions, namely, from the Poincar\'e-Miranda theorem, a useful but comparatively little known equivalent formulation of Brouwer's fixed point theorem.\footnote{See, e.g., \cite{browder1983} for background.} 

The connection between our model and models of interbank default networks such as \cite{eisenberg2001systemic} is, of course, not merely a formal, mathematical one. Depending on the choice of the fair conversion threshold, our model interpolates between two credit market models without CoCos. As conversion thresholds go to zero, the model converges to the Eisenberg-Noe model in which banks are forced out of the market in case of illiquidity. Conversely, as conversion thresholds go to infinity, the model converges to a situation in which default cycles are avoided by canceling all debt. CoCos may thus strike an interesting middle ground between these two extremes. 

Our main technical contribution is developing techniques for proving existence of equilibrium in the super-fair case. Here, the mapping from equilibrium stock prices to underlying asset values is piece-wise linear but discontinuous. Our basic strategy is to view the super-fair case as a distortion of a fair case with adjusted credit amounts. We provide an explicit fixed-point iteration that recovers an equilibrium of the super-fair case from equilibria of the fair case for different vectors of asset values, thus proving existence.  

\subsubsection*{Related Literature}

While this paper appears to be the first that studies CoCos with stock price triggers in a network setting, a number of papers have studied interaction effects in models with other types of trigger mechanisms.  In these models, ensuring well-definedness of equilibrium is simpler as there is no dependence of the stock price on itself, i.e., there are no circularity problems in the single bank case.

\cite{CW} consider CoCos with a regulatory trigger. In their setting only one of the banks issues CoCos and the focus is on the signaling value of conversions in an incomplete information model.\footnote{\cite{CW} are interested in whether CoCo conversions can trigger bank runs in a model as in \cite{DD}. As they regulatory triggers, conversion events reveal some of the regulator's inside information to the market. Interconnectedness in their model works only indirectly through an information externality that the CoCo issuer exerts on other banks. If all banks' returns are positively correlated, the conversion of one bank carries bad news about the returns of all banks in the market.}  In contrast, we consider complete information and study the market's ability to reflect all available information in prices and conversion events.

A number of  very recent papers study network effects of CoCos with accounting-based triggers, building on the single-bank model of \cite{GN12} \citep[rather than][]{GN}. Of these recent contributions, \cite{FH20} is closest in spirit to our paper, proving existence of equilibrium in a model with accounting-based triggers but allowing, e.g., for CoCos with different maturities. \cite{GWL20} show both in simulations and empirically that CoCos are an effective instrument for mitigating systemic risk. Beyond simple accounting-based triggers, they also consider extensions where conversion mechanisms take into account the balance sheets of the entire banking system.

From a broader perspective, our clear-cut existence and uniqueness results stand in interesting contrast to recent results on default in interbank networks with credit default swaps by \cite{schuldenzucker2017complexity,schuldenzucker2017default} where even the problem of deciding whether an equilibrium exists may be computationally intractable.\footnote{Algebraically, the problems studied in these works are quite different from ours. Our equilibrium conditions are piecewise linear and, in the super-fair case, discontinuous. The equations in \cite{schuldenzucker2017complexity,schuldenzucker2017default} are quadratic and thus ``more non-linear'' but continuous.}

\section{The Setting}\label{sec:set}

\subsubsection*{The Model}

We consider a multi-bank generalization of the static baseline model in \cite{SW} and \cite{GN}. Denote by $[n]=\{1,\dots,n\}$ the set of $n$ banks. For each bank $i\in [n]$, we denote by $a_i$ its assets net of liabilities. We assume that, in addition, banks have issued convertible debt. We denote by $c_i$ the total convertible debt that bank $i$ has to pay back. If bank $i$'s stock price $s_i$ turns out to be less than the conversion threshold $l_i$, the debt is converted (i.e. not paid back) and compensated by the issuing of $m_i>0$ new stocks. The original number of stocks is normalized to 1. Upon conversion, the former owners thus keep a fraction $1/(1+m_i)$ of bank $i$, while the former creditors receive a fraction $m_i/(1+m_i)$.

We depart from the previous literature by assuming that banks may have traded some of their convertible debt between each other. By $w_{ij}$, we denote the fraction of bank $j$'s convertible debt that is due to bank $i$. Depending on whether bank $j$ converts or not, bank $i$ thus either receives $w_{ij} m_j$ stocks or a cash amount of $w_{ij} c_j$. Consequently, the numbers $(w_{ij})$ can be interpreted as the adjacency matrix of a directed, weighted graph, the conversion network. The $w_{ij}$ satisfy $w_{ij}\in [0,1]$, $w_{ii}=0$ and $\sum_{i\in [n]} w_{ij} \leq 1$. The case $\sum_{i\in [n]} w_{ij} < 1$ corresponds to a setting where some of the convertible debt was issued to parties outside the banking system. In the degenerate case $w_{ij}=0$ for all $i$ and $j$, our model essentially collapses to $n$ independent copies of \cite{GN}'s model.

We consider a static model where all debt is settled simultaneously. What makes analyzing this setting challenging is the following dependence of the equilibrium prices $s=(s_1,\ldots,s_n)$ on themselves: Whether a bank $i$ converts, depends on whether its stock price $s_i$ is above or below $l_i$. Yet, whether the stock price is above or below $l_i$ depends on what it is worth to own the stock -- which depends itself on whether bank $i$ converts. This issue arises already in the previously studied single-bank setting. In our multi-bank setting, there is an additional layer of complexity as banks' conversions and stock prices are interrelated. The value of each bank depends on whether that bank receives stocks or cash from the other banks.

\subsubsection*{Equilibrium Concept}

Given  a partition $(B,C,H)$ of the set $[n]$ of banks into bankrupt ($B$), converting ($C$) and healthy ($H$) banks, we say that the price vector $s=(s_1,\ldots,s_n)$ and asset vector $a=(a_1,\ldots,a_n)$ form a $(B,C,H)$-equilibrium candidate if they solve the following system of equations:
\begin{align}\label{eqS1}
(1+m_i) s_i &= a_i + \sum_{j \in C} w_{ij}m_j s_j + \sum_{j \in H} w_{ij} c_j &\text{for all }i \in B \cup C \\
 s_i &= a_i-c_i + \sum_{j \in C} w_{ij}m_j s_j + \sum_{j \in H} w_{ij} c_j &\text{for all }i \in H
\label{eqS2}
\end{align}
Equation \eqref{eqS1} states that the total value of a converting bank's issued stocks $(1+m_i) s_i$ is equal to the assets $a_i$ plus the stocks the bank receives from converting banks plus the cash the bank receives from healthy banks. For a healthy bank as described in \eqref{eqS2}, two things are different: The total number of stocks on the left hand side is smaller and the debt $c_i$ is paid back on the right hand side. Bankrupt banks are also covered by equation \eqref{eqS1}. In their case, $s_i$ should not be interpreted strictly as the stock price but rather as a candidate for what the stock price would be if the bank was not bankrupt.\footnote{The true stock price of a bankrupt bank is, of course, zero. Note that for a bank $j\in B$, the value of $s_j$ does not appear on the right hand side of \eqref{eqS1} or \eqref{eqS2} for any other bank $i \neq j$. Moreover, if $s_i$ computed from \eqref{eqS1} is negative for a bank $i \in B$ then so is $s_i$ computed from \eqref{eqS2}. Thus, a negative $s_i$ from \eqref{eqS1} implies that neither converting nor being healthy are viable alternatives to bankruptcy.} For simplicity, we nevertheless call $s$ a vector of stock prices in the following.

A $(B,C,H)$-equilibrium candidate $(a,s)$ is a $(B,C,H)$-equilibrium if the partition $(B,C,H)$ is consistent with how the prices $s_i$ relate to the thresholds $l_i$: In equilibrium, we must have $B = \{i|s_i < 0\}$,  $C=\{i|0 \le s_i \le l_i\}$ and $H = \{i|s_i > l_i\}$.

As a first result, we show that for every stock price vector $s\in\mathbb{R}^n$ there exist a unique associated vector of asset values and a partition into bankrupt, converting and healthy banks.

\begin{lem}\label{lem:ex1}
For every $s\in \mathbb{R}^n$, there exist a unique partition $(B,C,H)$ of $[n]$ and an asset price vector $a$ such that $(a,s)$ is a $(B,C,H)$-equilibrium. Specifically, the partition $(B,C,H)$ is given by $B = \{i|s_i < 0\}$,  $C=\{i|0 \le s_i \le l_i\}$ and $H = \{i|s_i > l_i\}$ and the asset price vector is given by the unique solution $a$ to the linear system  (\ref{eqS1}--\ref{eqS2}) for this partition $(B,C,H)$.
\end{lem}

Unless otherwise noted, all proofs are in the appendix. The idea of the lemma is simply that the location of $s_i$ in relation to $0$ and $l_i$ determines whether bank $i$ should go bankrupt, convert or stay healthy. Yet once this information is available for all banks, we know the partition and computing $a$ from $s$ is reduced to solving the linear system (\ref{eqS1}--\ref{eqS2}). The lemma thus shows that there exists a mapping $\Phi:\mathbb{R}^n \rightarrow \mathbb{R}^n$ which maps stock price vectors to the unique asset value vectors that rationalize them in equilibrium.

The mapping $\Phi$ from stock prices to asset values is thus easy to understand and compute. Unfortunately, what matters in practice is the inverse of this mapping -- from asset values to stock prices. Given a realized vector of asset values, does there exist a unique stock price vector $s$ such that $\Phi(s)=a$? If this is not the case, there may be multiple candidates for equilibrium stock prices or there may be non-existence of equilibrium. The rationale behind this question is as follows: We consider a one period market model. In the first step, asset values (or asset minus liability values) of all banks realize to some $a\in \mathbb{R}^n$. In the next step, the market wishes to arrive at equilibrium stock prices for the banks. If a unique equilibrium exists, this is what the market will find eventually. If multiple equilibria exist, the market is confused. If no equilibrium exists, the market is unpredictable.

Inspecting only (\ref{eqS1}--\ref{eqS2}), it may seem at first sight that the relation between $a$ and $s$ is linear. Yet in fact, it is piece-wise linear due to the partitions $(B,C,H)$. The equilibrium partition can easily be read-off from $s$ but not from $a$. This makes the mapping from $a$ to $s$ more difficult to analyze than the mapping from $s$ to $a$. The intuition for this asymmetry is straightforward. Whether bank $i$ is healthy, converting or bankrupt only depends on the stock price $s_i$. Yet, due to interconnectedness, the stock price $s_i$ may depend on the asset values of many banks.

\subsubsection*{Rewriting the problem}

So far, we have seen that questions of existence and uniqueness of equilibrium can be reduced to structural properties of the mapping $\Phi$. Existence of equilibrium holds if the mapping $\Phi$ is a surjection from $\mathbb{R}^n$ to $\mathbb{R}^n$, i.e., if for every asset value $a\in \mathbb{R}^n$ there exists a stock price vector $s \in \mathbb{R}^n$ such that $a=\Phi(s)$. Surjectivity by itself does not imply uniqueness of equilibrium, i.e., under surjectivity $a$ could be the unique rationalizing asset value vector for more than one stock price vector.  We have existence and uniqueness of equilibrium if the mapping $\Phi$ is a bijection from $\mathbb{R}^n$ to $\mathbb{R}^n$, i.e., if for every $a$ there exists a unique $s$ such that $a=\Phi(s)$.

In order to understand the mapping $\Phi$ better, we need to introduce some additional notation. Given a vector $v = (v_1,\dots,v_n)$ and a set $F \subset [n]$ we denote by $v_F\in\mathbb{R}^n$ the vector defined by $(v_F)_i = v_i$ for all $i \in F$ and $v_i =0$ for all $i \notin F$. For any vector $v\in \mathbb{R}^n$ we denote by $\Diag(v)$ the diagonal matrix in $\mathbb{R}^{n \times n}$ with diagonal entries $v$. We denote by $e = (1,\dots,1)\in \mathbb{R}^n$ the all ones vector and by  $e_i = e_{\{i\}}$ the vector that is all zeros except for a 1 in position $i$. For a given partition $(B,C,H)$, conditions \eqref{eqS1} and \eqref{eqS2} can be summarized as
\begin{align}\label{eqS3}
a = s + \Diag(m_B) s + (I- W) \Diag(m_C) s + (I-  W) c_H.
\end{align}
Thus, for a given partition $(B,C,H)$, $a$ is an affine function of $s$ and we have $a = L_{B,C}s + b_{H}$ where $L_{B,C} = I + \Diag(m_B)  + (I- W) \Diag(m_C) $ and $b_{H} =  (I-  W) c_H$. Due to the fact that $\sum_{i\in [n]} w_{ij} \leq 1$ and $m_j > 0$, the matrices $L_{B,C}$ are strictly diagonally dominant and thus of full rank and invertible. This shows that for a given partition and a given  $s$ the vector $a$ is uniquely determined.

\section{Equilibrium Analysis}

\subsection{Overview}\label{sec:overview}
One main observation so far is that any vector of stock prices determines a unique associated partition of bankrupt, converting and healthy banks. For a given partition $(B,C,H)$, denote by $\cS_{B,C,H}$ the set of stock price vectors $s$ which lead to this partition, i.e., $\cS_{B,C,H} = \{s| s_i <0 \text{ for } i \in B,\, 0 \le s_i \le l_i \text{ for } i \in C \text{ and } s_i > l_i \text{ for } i \in H \}$. Clearly,
the sets $\cS_{B,C,H}$ form a partition of $\mathbb{R}^n$ into $3^n$ disjoint sets. We denote by $\cA_{B,C,H}$ the image of $\cS_{B,C,H}$ under $\Phi$, i.e.,
\begin{align}\label{ABCH}
\cA_{B,C,H} =\Phi(\cS_{B,C,H})= \{L_{B,C}s + b_{H}|s \in \cS_{B,C,H}\}
\end{align}
since, by the previous discussion, $\Phi$ is defined as
$\Phi(s)=L_{B,C}s + b_{H}$ for $s\in \cS_{B,C,H}$. To understand whether the mapping $\Phi$ is surjective or even bijective, we need to understand conditions under which the union of the sets $\cA_{B,C,H}$ over all possible partitions covers the entire space $\mathbb{R}^n$ and under which this covering is disjoint. As a starting point, notice that the sets  $\cS_{B,C,H}$  are polyhedra and that the sets $\cA_{B,C,H}$ are affine transformations of polyhedra -- and thus also polyhedra. The reason is that the restriction of $\Phi$ to $\cS_{B,C,H}$ is an affine function as shown in \eqref{ABCH}.

\begin{figure}[htb]
	\begin{center}
	\end{center}
	\begin{center}
		\includegraphics[width=0.3\textwidth]{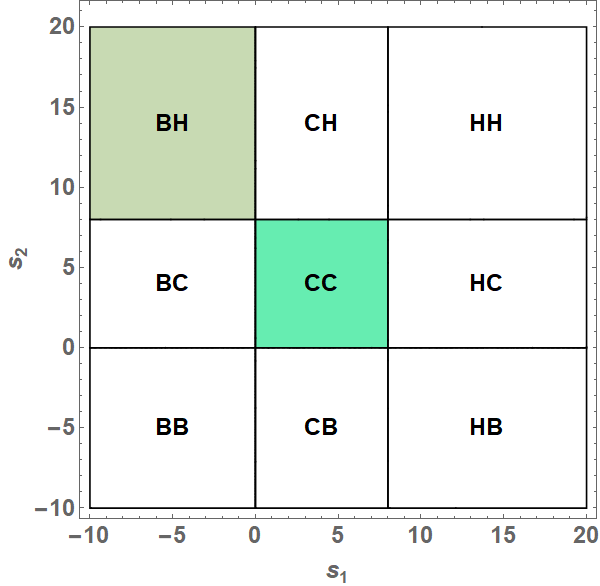}		
		\includegraphics[width=0.3\textwidth]{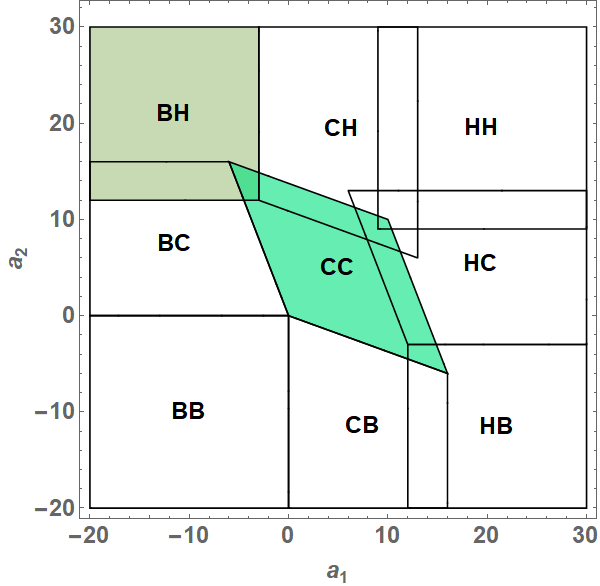}
\end{center}
	\caption{The mapping $\Phi$.}\label{figPhi}
	\vspace*{0.15cm}
	\parbox{\textwidth}{\footnotesize The parameters are $m_1=m_2=1, c_1=c_2=4$, $w_{12}=w_{21}=0.75$ and $l_1=l_2=8$. The stock price space is shown on the left and the asset value space is shown on the right. }
\end{figure}

For the case $n=2$, Figure \ref{figPhi} shows the sets $\cS_{B,C,H}$ (left panel) and the associated sets $\cA_{B,C,H}=\Phi(\cA_{B,C,H})$ (right panel) in the respective spaces of stock prices and asset values. Highlighted are the sets where both banks convert (CC) in the middle, and the set where bank 1 is bankrupt and bank 2 is healthy in the upper left corner (BH). In terms of partitions, the two sets correspond to, respectively, $(B,C,H)=(\emptyset,\{1,2\},\emptyset )$ and $(B,C,H)=(\{1\},\emptyset,\{2\})$. We see that the two highlighted sets $\cA_{B,C,H}$ are overlapping, indicating that for this parameter constellation the mapping $\Phi$ is (surjective but) not bijective. There thus exist combinations of asset values that may give rise to more than one equilibrium partition and thus more than one vector of equilibrium stock prices.

One main contribution of our paper is to formulate precise conditions for existence and uniqueness of equilibrium in this model in terms of the conversion thresholds chosen by different banks. For instance, the situation depicted in Figure \ref{figPhi} is one of existence and non-uniqueness. In the remainder of this section, we illustrate these conditions for the two bank case and introduce the necessary terminology. The formal existence and uniqueness results follow in later sections.

As a first step, compare the two candidates $s_i^c$ and $s_i^h$ for the stock price of bank $i$ in case of conversion and non-conversion
implied by  (\ref{eqS1}--\ref{eqS2}),
\begin{align} s_i^c &= a_i -m_i s_i^c+ \sum_{j \in C} w_{ij}m_j s_j + \sum_{j \in H} w_{ij} c_j,\nonumber\\
 s_i^h &= a_i\;-\;c_i\;\; + \sum_{j \in C} w_{ij}m_j s_j + \sum_{j \in H} w_{ij} c_j.\nonumber
\end{align}
The two candidate stock prices differ only in the transfer $c_i$ vs. $m_i s_i^c$ that is made to the creditors of bank $i$. We say that the threshold set by bank $i$ is \textit{fair} if creditors are indifferent between conversion and non-conversion for marginal cases, i.e., for conversions at the threshold $s_i=l_i$. The fair threshold for bank $i$ is thus given by $c_i=m_i l_i$, i.e., $l_i=c_i/m_i$. We say that the threshold set by bank $i$ is super-fair if $l_i> c_i/m_i$, i.e., if conversions at the threshold correspond to a wealth transfer from the bank's original stock holders to the creditors. Conversely, we call thresholds  $l_i$ with $l_i< c_i/m_i$ sub-fair.

\begin{figure}[htb]
	\begin{center}
	\end{center}
	\begin{center}
		\begin{subfigure}[b]{0.3\textwidth}
			\includegraphics[width=\textwidth]{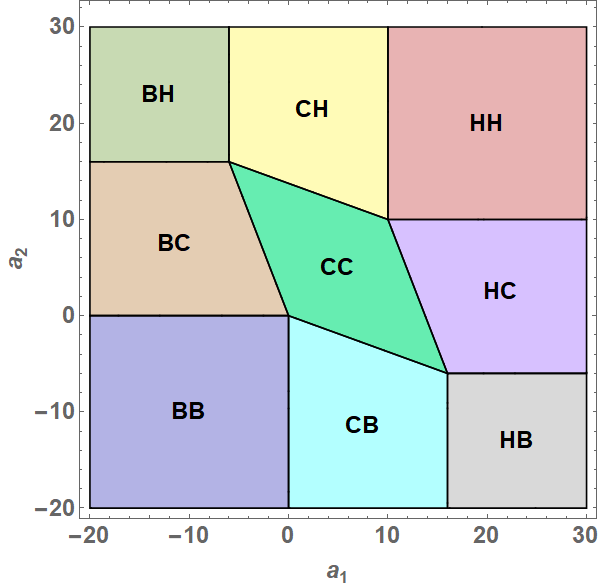}
			\caption{Fair}\label{fig2:a}
		\end{subfigure}
	\begin{subfigure}[b]{0.3\textwidth}
		\includegraphics[width=\textwidth]{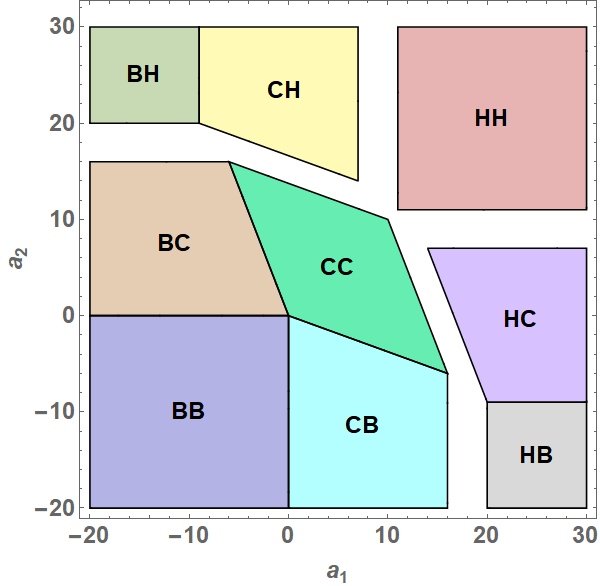}
		\caption{Sub-fair}\label{fig2:b}
	\end{subfigure}
	\begin{subfigure}[b]{0.3\textwidth}
		\includegraphics[width=\textwidth]{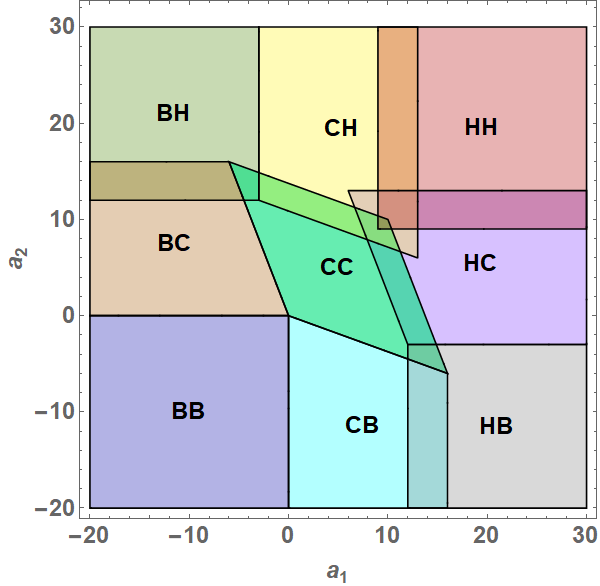}
		\caption{Super-fair}\label{fig2:c}
	\end{subfigure}
	\end{center}
	\caption{Non-existence, uniqueness and multiplicity of equilibria.}\label{fig2}
		\vspace*{0.15cm}
	\parbox{\textwidth}{\footnotesize The parameters are $m_1=m_2=1$, $w_{12}=w_{21}=0.75$ and $l_1=l_2=8$. Credit amounts are $(c_1,c_2)=(8,8)$ in (a),  $(c_1,c_2)=(12,12)$ in (b) and $(c_1,c_2)=(4,4)$ in (c). }
\end{figure}

In the numerical example above, fair thresholds are given by $l_i=c_i/m_i=8$ so that the situation of equilibrium existence and non-uniqueness depicted in Figure \ref{figPhi} corresponds to an example of the super-fair case. A first illustration of the more general picture is given in Figure \ref{fig2} which shows the partition in asset value space for different choices of issued debt. 
The left panel, corresponding to fair thresholds, shows a non-overlapping partition of the asset value space and thus a
situation of existence and uniqueness of equilibrium. The middle panel, corresponding to sub-fair thresholds by both banks, shows a non-overlapping partition with gaps. Thus, whenever equilibrium exists it is unique. However, there exist combinations of asset values which do not lie in any of the sets $\cA_{B,C,H}$. For these asset values, no equilibrium stock prices exist. The right panel, corresponding to super-fair thresholds by both banks, shows an overlapping partition. In this case, every constellation of asset values leads to at least one vector of equilibrium stock prices, but in some cases, we see an overlap corresponding to multiple equilibria. For example, we see areas where (only) the sets HC and CH overlap. Here, one bank has to convert in equilibrium to save the other -- but it is not determined which bank that is.

We close this section with a preview of our main results for the $n$ bank case and a classification into fair, super-fair and sub-fair markets.
\begin{itemize}
\item If all banks set fair thresholds, there exists a unique equilibrium for any vector of asset values. We call this the fair case.
\item If all banks set fair or super-fair thresholds, there exists an equilibrium stock price for any vector of asset values. We call this the super-fair case.
\item If some bank sets a sub-fair threshold, there exists a vector of asset values for which no equilibrium stock price exists. We call this the sub-fair case.
\end{itemize}
Note that the sub-fair case is defined a bit more broadly: Suppose some banks set fair thresholds, some set super-fair thresholds while others set (strictly) sub-fair thresholds. Then we say we are in the sub-fair case because we have non-existence of equilibrium which is more severe than the non-uniqueness implied by the super-fair thresholds. This is our first main result.

\begin{prop}\label{prop:subfair}
Suppose we are in the sub-fair case, i.e., there exists a bank $i$ with $l_i<c_i/m_i$. Then there exists a vector $a\in\mathbb{R}^n$ such that $\Phi(s)\neq a$ for all $s\in \mathbb{R}^n$.
\end{prop}

The intuition behind the proof is straightforward. When the asset values of all banks except for $i$ are sufficiently high above the conversion thresholds, then the possible conversion of bank $i$ can be studied in isolation as all other banks can only be healthy in equilibrium. Non-existence then follows by the same argument as in the single bank model of  \cite{SW} and \cite{GN}.

\begin{figure}[htb]
	\begin{center}
	\end{center}
\begin{center}
		\begin{subfigure}[b]{0.3\textwidth}
	\includegraphics[width=\textwidth]{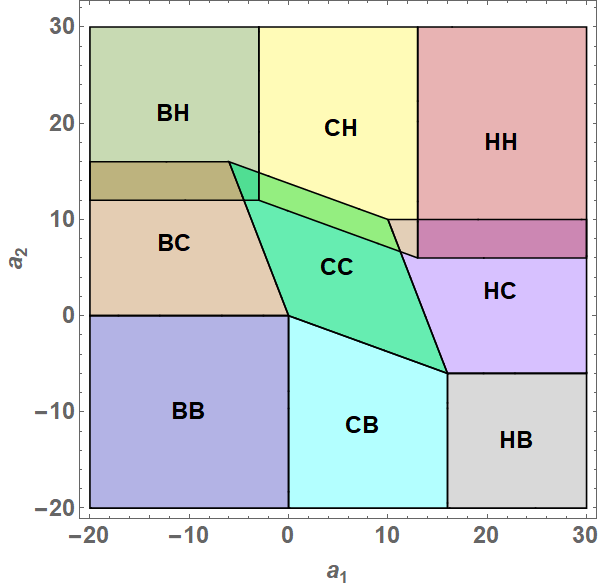}
	\caption{Fair / super-fair}\label{fig3:a}
\end{subfigure}
\begin{subfigure}[b]{0.3\textwidth}
	\includegraphics[width=\textwidth]{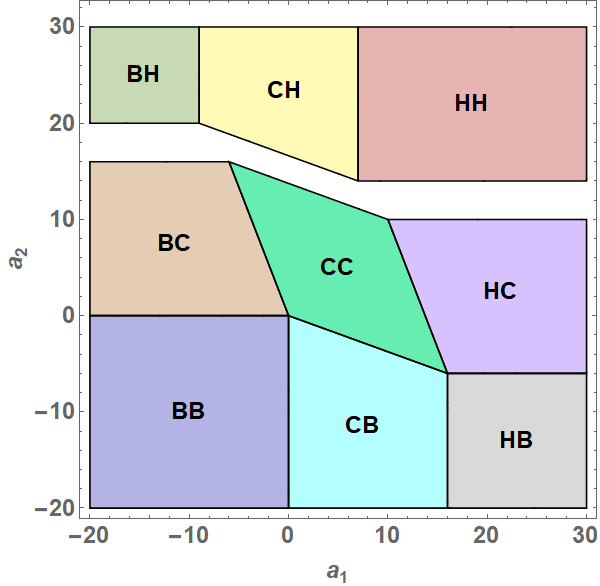}
	\caption{Fair / sub-fair}\label{fig3:b}
\end{subfigure}
\begin{subfigure}[b]{0.3\textwidth}
	\includegraphics[width=\textwidth]{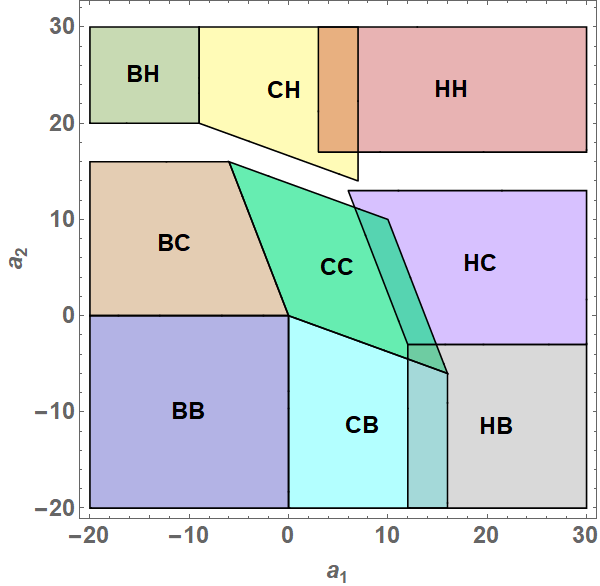}
	\caption{Super-fair / sub-fair}\label{fig3:c}
\end{subfigure}
\end{center}
\caption{Asymmetric examples of non-existence and multiplicity of equilibria.}\label{fig3}
\vspace*{0.15cm}
\parbox{\textwidth}{\footnotesize The parameters are $m_1=m_2=1$, $w_{12}=w_{21}=0.75$ and $l_1=l_2=8$. Credit amounts are $(c_1,c_2)=(8,4)$ in (a),  $(c_1,c_2)=(8,12)$ in (b) and $(c_1,c_2)=(4,12)$ in (c). }
\end{figure}

Figure \ref{fig3} shows the partition in asset value space for asymmetric choices of issued debt. In the left two panels, bank 1 is in a fair case while bank 2 is, respectively, super-fair and sub-fair. In the right panel, bank 1 is super-fair while bank 2 is sub-fair. We see that, essentially, the choices of one bank cannot cure the problems that may arise due to the choices of the other bank. There are also immediate spillovers even if only one bank has a non-fair threshold. For instance, in the left panel we observe an overlap between the sets denoted CC and BH, corresponding to a situation with two possible equilibria. Either both banks convert or bank 1 goes bankrupt while bank 2 is healthy. Thus, in one equilibrium, the second bank converts to save the first bank. In the other equilibrium this is not the case.

\subsection{Existence and uniqueness of equilibrium in the fair case}\label{sec:fair}

In the next step, we show that in the fair case, i.e., when the thresholds are given by $l_i=c_i/m_i$ for all $i$, the mapping $\Phi$ from stock prices to asset values is surjective. This implies that for every vector of asset values $a$ we can find a vector of stock prices $s$ such that $a$ is the unique vector of asset values which rationalizes $s$. Our existence proof relies on a multivariate version of the intermediate value theorem which is given next. The result is a corollary of the Poincar\'e-Miranda theorem which  is itself an equivalent formulation of Brouwer's fixed point theorem.

\begin{prop}\label{propIVT}
Let $f:\mathbb{R}^n \rightarrow \mathbb{R}^n$,  $f(x)=(f_1(x_1,\ldots,x_n ),\ldots, f_n(x_1,\ldots,x_n ))$, be a continuous function with the following properties:
\begin{itemize}
\item[(i)] For all $i\in [n]$, $f_i(x)$ is weakly increasing in $x_i$ and weakly decreasing in $x_j$, $j\neq i$.
\item[(ii)] For all $i\in [n]$, we have $\lim_{t\rightarrow \infty} f_i(t e)=\infty$ and $\lim_{t\rightarrow \infty} f_i(- te) =- \infty$. 
\end{itemize}
Then $f$ is surjective, i.e., for every $y\in \mathbb{R}^n$ there exists $x \in \mathbb{R}^n$ such that $f(x)=y$.
\end{prop}

The proposition is a multivariate version of the observation that a continuous, univariate function $f(x)$ that goes to $\pm \infty$ as $x$ goes to $\pm \infty$ must pass through every point. To guarantee that a continuous, multivariate function passes through every vector, we need to assume some more structure however. In (i), we assume monotonicity in all components. In (ii), we assume that, essentially, when all components of $x$ go to $\pm \infty$ then so do all components of $f$.\footnote{The formulation of the proposition assumes that $f_i$ increases in $x_i$ and decreases in $x_j$, $j \neq i$ which is what we need here. Condition (i) can easily be relaxed to functions whose components are either increasing or decreasing in all components when condition (ii) is suitably adapted.} In our setting of convertible debt, this condition holds because the asset values of other banks become irrelevant for the status of bank $i$ if its own assets are sufficiently positive or negative. The reason is simply that borrowed amounts $c_{j}$ are finite and that conversions only happen at intermediate stock price values.

We apply Proposition \ref{propIVT} in our setting by verifying that in the fair case the function $\Phi$ satisfies all of its requirements. Most of the work here comes from verifying that $\Phi$ is continuous. Once this is shown, the monotonicity properties (i) and (ii) follow easily as $\Phi$ is a locally linear function.

\begin{prop}\label{prop:surj}
In the fair case with  $l_i=c_i/m_i$ for all $i \in [n]$, the mapping $\Phi$ is surjective and continuous. Thus, for every asset value vector $a\in\mathbb{R}^n$
there exists a partition $(B,C,H)$ of $[n]$ and a stock price vector $s$ such that $(a,s)$ is a $(B,C,H)$-equilibrium.
\end{prop}

To conclude our analysis of the fair case, we show that the mapping $\Phi$ is not only surjective but also injective. The main difficulty here is that the components of $\Phi$ are increasing in one coordinate but decreasing in the others. Thus, one might be worried that movements in different directions could cancel each other out. We rely on the strict diagonal dominance of the matrices  $L_{B,C}$ to argue that this cannot be the case. 

\begin{prop}\label{prop:inj}
	In the fair case with  $l_i=c_i/m_i$ for all $i \in [n]$, the mapping $\Phi$ is injective, and thus also bijective. Thus, for every asset value vector $a$ 	there exists a unique partition $(B,C,H)$ of $[n]$ and a stock price vector $s$ such that $(a,s)$ is a $(B,C,H)$-equilibrium.
\end{prop}

This bijectivity result proves existence and uniqueness of equilibrium: For any vector of asset values $a$ there exists a unique matching vector of equilibrium stock prices. 

\subsection{Existence of equilibrium in the super-fair case}\label{sec:superfair}
In this section, we show that there always exists an equilibrium when all banks have set fair or super-fair thresholds, $l_i \geq c_i/m_i$ for all $i\in [n]$. Throughout this section, we impose one further technical condition, assuming invertibility of the matrix $I-W$. 
\begin{assumption}\label{A1}
The matrix $I-W$ is invertible.
\end{assumption}
Intuitively, Assumption \ref{A1} means that at least a tiny fraction of CoCos has been sold to parties outside the banking system.\footnote{We conjecture that this assumption can be removed at the expense of more technical proofs.} The main difficulty in the proof is that the mapping from stock prices to asset values is no longer continuous as in the fair case. Thus, there is little hope for proving surjectivity based on variations of Brouwer's fixed point theorem. Instead, our basic strategy is to view the super-fair case as a distortion of the fair case. We argue that, unlike continuity and injectivity, surjectivity of the mapping from stock prices to asset values is preserved under this distortion. This implies that for every vector of asset value there is at least one associated vector of stock prices. 

In the fair case, the mapping $\Phi$ is given by $\Phi(s) = L_{B,C}s + b_H$ where $L_{B,C} = I + \Diag(m_B)  + (I- W) \Diag(m_C) $ and $b_H=(I-W)c_H$ for $s\in \mathcal{S}_{B,C,H}$. 
Fairness means that $l_j = c_j/m_j$ for all $j\in [n]$.  In the fair case, $\Phi$ is a bijection. Thus we can define $H(a) = \{i:\Phi^{-1}(a)_i > \ell_i\} $, the set of banks which are healthy in $a$ in the fair case.

Now we introduce the super-fair case $\hat \Phi$ which has the same values of $m$, $W$ and $l$ but smaller credit amounts, $\hat{c}_j = c_j -d_j$, $d_j \geq 0$ for all $j\in [n]$. We keep the vector $d \in \R^n_+$ fixed throughout this section.  Since any super-fair case can be written as a distorted fair case with decreased credit amounts, it suffices to show that $\hat \Phi$ is surjective.

For  $s\in \mathcal{S}_{B,C,H}$, we know that $\hat \Phi(s) = L_{B,C}s + (I-W)\hat{c}_H$
which implies that the two mappings $\Phi$ and $\hat{\Phi}$ are related via $$\hat \Phi(s) = \Phi(s) - (I-W)d_{H(\Phi(s))}$$ for all $s$. Thus, the difference between $\Phi$ and $\hat{\Phi}$ only depends on the set of healthy banks $H(a)$. Moreover, by the relationship
\[
(I-W)d_{H(a)} = \sum_{j\in H(a)} (I-W)d_{\{j\}}=\sum_{j\in H(a)} d_j(I-W)e_j,
\]
we see that effectively, every bank $j$ has a shift vector $(I-W)d_{\{j\}}$ that it contributes to the distortion from the fair to the super-fair case whenever it is healthy in  $a$. Thus, in the two bank case, an asset value at which only bank 1 is healthy is shifted by $(I-W)d_{\{1\}}$ while an asset value at which both banks are healthy is shifted by both $(I-W)d_{\{1\}}$ and $(I-W)d_{\{2\}}$. It is critical to understand where such combinations of shifts can lead us. 

\begin{figure}[htb]
	\begin{center}
			\includegraphics[width=0.3\textwidth]{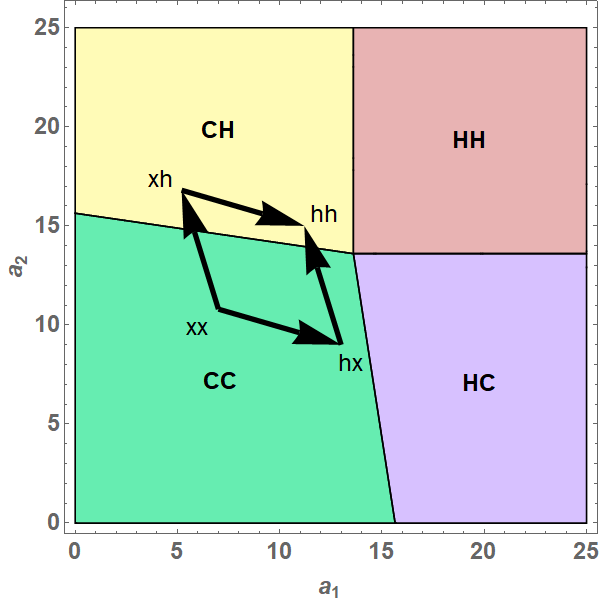}
			\includegraphics[width=0.3\textwidth]{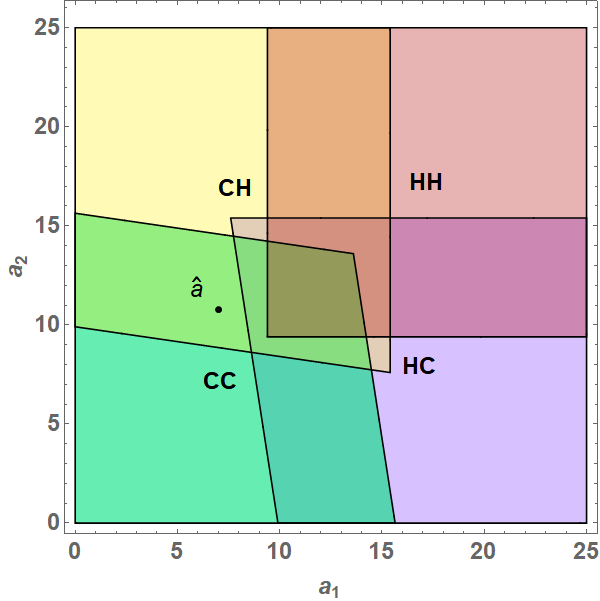}\\
			\includegraphics[width=0.3\textwidth]{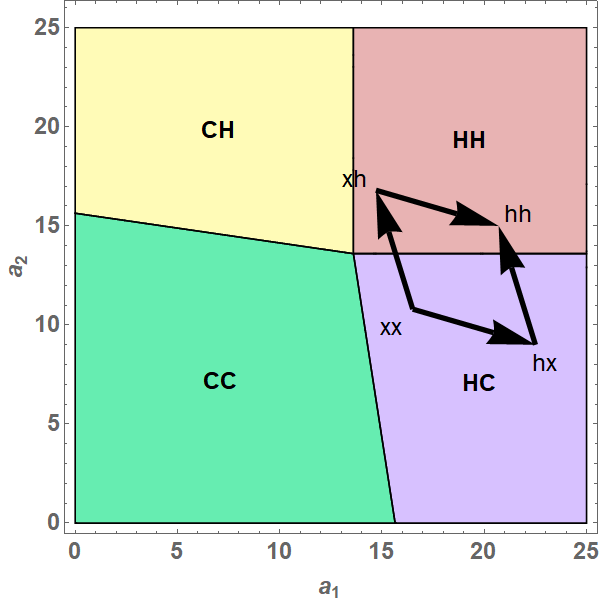}
			\includegraphics[width=0.3\textwidth]{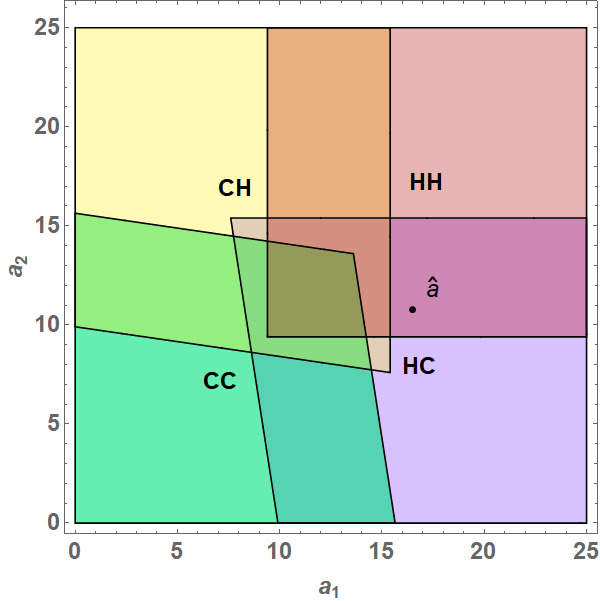}
	\end{center}
	\caption{Strategy of proof.}\label{fig_proof}
	\vspace*{0.15cm}
	\parbox{\textwidth}{\footnotesize 
		In all four panels, we have  $m_1=m_2=1$, $l_1=l_2=8$ and $w_{12}=w_{21}=0.3$. In the left panels, we have fair credit amounts, $c_1=c_2=8$, while in the right panels, we have $\hat{c}_1=\hat{c}_2=2$. The shift vectors are given by $(I-W)d_{\{1\}}=(I-W)(c-\hat{c})_{\{1\}}=(6,-1.8)$ and $(I-W)d_{\{2\}}=(-1.8,6)$.}
\end{figure}

Figure \ref{fig_proof} illustrates the central ideas of our surjectivity proof for two points $\hat{a}$. For convenience, we only plot the positive quadrant. The two right panels show an instance of the super-fair case while the left panels show the corresponding fair case with adjusted credit amounts. In the upper panels we have a situation where the possible equilibria in $\hat{a}$ in the super-fair case are either both banks converting or bank 1 converting while bank 2 is healthy. In the lower panels, the two equilibria are either both being healthy or bank 1 being healthy while bank 2 converts. We can easily see this by studying the overlaps in the right panels. The key observation is that we can also see this by studying the trapezoid spanned by the arrows in the left panel. Here, we have drawn the four points that can be reached by applying the shift vectors to $\hat{a}$. $hh$ is the point we reach when we apply both vectors, $hx$ is the point we reach when we apply the first shift vector and so on. Here ``$x$'' stands for ``$b$'' or ``$c$'', bankrupt or converting, and thus no shift.  

Our claim is that we can find the equilibria of the super-fair case by studying which of the corners of the trapezoid lie in matching sets of the partition. In the upper panels, the point $xx$ lies in CC and the point $xh$ lies in CH. These we count as matches because the same banks are healthy. The points $hx$ and $hh$ lie in CC and CH so these are not matches. Indeed, CC and CH correspond to the two equilibria at $\hat{a}$ in the super-fair case. Similarly, in the lower panels we count matches for the points $hx$ and $hh$ which corresponds to the equilibria in HC and HH. 

Thus, in this two-bank example, proving surjectivity boils down to showing that no matter where we place the trapezoid in the picture, one corner will always be in the matching set. The general version of this claim is formalized as a fixed point problem in Lemma \ref{lem:fixpoint}, the main technical result of this section. For every point $a$ there exists a set $X$ of banks with the following property. The set of banks which are healthy in the point $a+(I-W)d_X$ is $X$ itself. Here, the point $a+(I-W)d_X$ is the point we reach when applying the shift vectors associated with $X$ to the starting point $a$.

\begin{lem}\label{lem:fixpoint} Under Assumption \ref{A1}, for all $a \in \R^n$ there is $X \subset [n]$ such that $H(a+(I-W)d_X) = X$.
\end{lem}

Once we have established Lemma~\ref{lem:fixpoint}, the surjectivity of $\hat \Phi$ is easy to show. This is the content of the following theorem. Intuitively, its short proof establishes our claim about the connection between the left and right panels in Figure \ref{fig_proof}.

\begin{thm}\label{thm:phihat}
Under Assumption \ref{A1},  the mapping $\hat \Phi$ associated with the super-fair case is surjective. Thus, for all vectors of asset values $a\in \R^n$ there exists an associated vector of equilibrium stock prices $s$ with $\hat{\Phi}(s)=a$.
\end{thm}
\begin{myproof}{Theorem}{\ref{thm:phihat}}
  Let $a\in \R^n$ be given. From Lemma~\ref{lem:fixpoint} there is $X \subseteq [n]$ such that $H(a+(I-W)d_X) = X$. From the definition of $\hat \Phi$ we obtain $\hat\Phi(\Phi^{-1}(a+(I-W)d_X)) = a +(I-W)d_X - (I-W)d_{H(a+(I-W)d_X)} = a$.
\end{myproof}

As a first step towards the proof of Lemma \ref{lem:fixpoint}, we prove two qualitative results about the interaction between shifts and sets of healthy banks. The first lemma considers banks that are healthy after shifting the banks in a set $X$. The claim is that all these banks lie in the union of the shifted banks, $X$, and the banks that were originally healthy, $H(a)$. In the second lemma, we consider shifting some banks which are healthy in $a$. The claim is that after this shift some of the shifted banks must still be healthy.

\begin{lem}\label{lem:preserveSomeHealth} Let $a \in \R^n$ be given. For any $X \subseteq [n]$,
$H(a+(I-W)d_X) \subseteq H(a) \cup X$.
\end{lem}

\begin{lem}\label{lem:presSomeHealth3} Let $a \in \R^n$ be given. 
Then, under Assumption \ref{A1}, for any $X \subseteq [n]$, $\emptyset \neq X \subseteq H(a)$ implies $X \cap  H(a+(I-W)d_X) \neq \emptyset$.
\end{lem}

Essentially, our proof of Lemma \ref{lem:fixpoint} is now based on constructing a suitable sequence of subsets of $[n]$ and to show that is it contracting around a fixed point. To this end, for the remainder of this section, we  fix some  $\hat a \in \R^n$ and define the mapping $h(X) = H(\hat a + (I-W)d_X)$, $X  \subset [n]$. Thus, $h$ maps subsets of $[n]$ to subsets of $[n]$. What we need to show is that there exists a fixed point, i.e., an $X$ such that $h(X)=X$. We will construct such a fixed point by iterating the mapping $h$. 
As customary, given $X  \subset [n]$ we define $h^0(X) = X$ and $h^{m+1}(X) = h(h^m(X))$ for all $m \ge 0$.  In Corollary \ref{cor:antimon1}, we collect some facts about the monotonicity behavior of $h$. We see that $h$ is far from monotonic but rather exhibits some type of cyclical behavior when it is applied multiple times.

\begin{cor}\label{cor:antimon1}  Under Assumption \ref{A1}, let $X \subseteq [n]$.
\begin{enumerate}[(i)]
  \item\label{loc:antimon1} If $X \subseteq Y \subseteq  h(X)$ then $ h(Y) \subseteq h(X)$.
  \item\label{loc:antimon2} If $ h(X) \subseteq Y \subseteq X$ then $h(X) \subseteq h(Y)$.
  \item\label{loc:antimon3} If $ X \subseteq h^2(X) \varsubsetneq h(X)$ then $h^2(X) \subseteq h^3(X)\varsubsetneq h(X)$.
  \item\label{loc:antimon4} If $ h(X) \varsubsetneq h^2(X)\subseteq X$ then $h(X) \varsubsetneq h^3(X)\subseteq h^2(X)$.
\end{enumerate}
\end{cor}

Our strategy of proof for Lemma \ref{lem:fixpoint} is based on the sequence of sets given by $X_0 = H(\hat a)$ and $X_{m+1}=h(X_m)$ for $m\geq 0$. From the corollary, we can deduce that $X_1$ is a subset of $X_0$, that $X_2$ lies between $X_1$ and $X_0$, $X_3$ between $X_1$ and $X_2$ and so on. The fact that some of the claims in Corollary \ref{cor:antimon1} are strict inclusions, $\varsubsetneq$, then implies that this sequence of sets cannot cycle on forever but must become constant after finitely many steps, $h(X_m)=X_{m+1}=X_m$. This is the desired fixed point. 

\begin{figure}[htb]
	\begin{center}
		\includegraphics[width=0.3\textwidth]{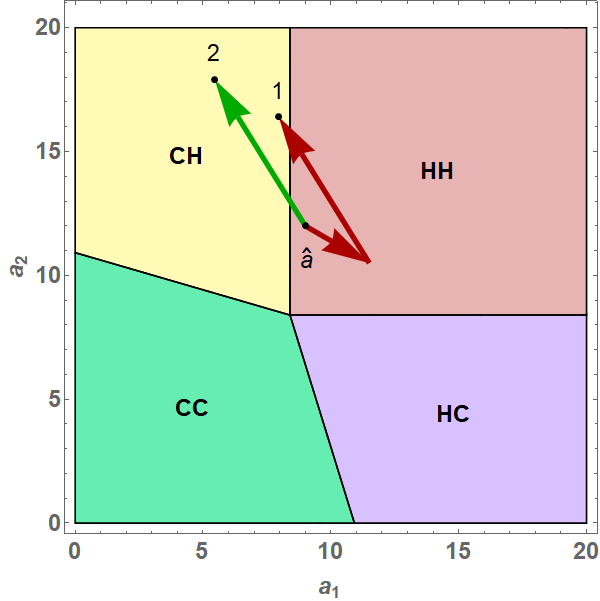}
	\end{center}
	\caption{Construction of fixed points.}\label{fig_proof2}
	\vspace*{0.15cm}
	\parbox{\textwidth}{\footnotesize 
	Equilibria for the fair case with $m_1=m_2=1$, $l_1=l_2=6$, $w_{12}=w_{21}=0.6$, $c_1=c_2=6$. The shift vectors correspond to the super-fair case with $\hat{c}_1=3.5$, $\hat{c}_2=0.1$ and thus $(I-W)d_{\{1\}}=(2.5, -1.5)$ and $(I-W)d_{\{2\}}=(-3.54, 5.9)$.}
\end{figure}

Figure \ref{fig_proof2} illustrates this construction. We start with a point $\hat{a}$ in which both banks are healthy, $X_0=H(\hat{a})=\{1,2\}$. The next step is thus to consider the healthy banks in the point $\hat{a}+(I-W)d_{X_0}$, shifting both banks. As can be seen from the point ``1'', this leads us into the set where only bank 2 is healthy, $X_1=h(X_0)=H(\hat{a}+(I-W)d_{X_0})=\{2\}$. Next, we consider $X_2=h(X_1)$, the healthy banks in the point $\hat{a}+(I-W)d_{X_1}$. From the point ``2'' in the figure, we see that $X_2=\{2\}=X_1$, the desired fixed point.   

\section{Conversions vs. Bankruptcies}\label{sec:disc}

In this section, we discuss how our model with CoCos can be seen as interpolating between two rather different scenarios without CoCos: As one extremal case, we obtain a model in the spirit of \cite{eisenberg2001systemic} where a bank's assets are fully used to settle outstanding debt in case of insolvency. At the other extreme, we find a model where effectively all debt is canceled.

As a first step, we rewrite our original model which is based on banks' stock prices $s_i$ into a model which is based on their total equity values $v_i$. The relation between stock prices and equity values is as follows. When bank $i$ is healthy, $i \in H$, we have $v_i=s_i$ as all equity belongs to the original stockholders. When bank $i$ is not healthy, $i \in B \cup C$, $m_i$ additional stocks are issued in the conversion process and we have $v_i=(1+m_i)s_i$.  

When we rewrite our previous equilibrium conditions in terms of total equity values, we obtain the following. Given  a partition $(B,C,H)$ of the set $[n]$ of banks into bankrupt ($B$), converting ($C$) and healthy ($H$) banks, we say that the vector of equity values $v=(v_1,\ldots,v_n)$ and asset vector $a=(a_1,\ldots,a_n)$ form a $(B,C,H)$-equilibrium candidate if they solve the following system of equations:
\begin{align}\label{eqE1}
 v_i &= a_i + \sum_{j \in C} w_{ij} \frac{m_j}{1+m_j} v_j + \sum_{j \in H} w_{ij} c_j &\text{for all }i \in B \cup C \\
 v_i &= a_i-c_i + \sum_{j \in C} w_{ij} \frac{m_j}{1+m_j} v_j + \sum_{j \in H} w_{ij} c_j &\text{for all }i \in H
\label{eqE2}
\end{align}
A $(B,C,H)$-equilibrium candidate $(a,v)$ is a $(B,C,H)$-equilibrium if $v_i<0$ for all $i\in B$, if $0 \le v_i \le (1+m_i) l_i$ for all $i \in C$ and if $v_i> l_i$ for all $i \in H$.\footnote{As before,  negative equity values in case of a bankruptcy should be read as theoretical candidate equity values. The actual equity values are, of course, zero.} Comparing \eqref{eqE1} and \eqref{eqE2}, we see that, around the conversion thresholds, total equity value is higher in the converting than in the healthy case. The reason is that in the conversion process some debt is canceled in exchange for a shift in the ownership structure which is not reflected in the total equity value.

Some properties of the system become clearer in this formulation than in our previous one based on $s$. First, the way the weights $m_i/(1+m_i)$ appear in \eqref{eqE1} and \eqref{eqE2} suggests that by decreasing $m_i$ we reduce the strength of interactions between banks. We discuss this in more detail below. Second, unlike our previous comparison of $s_i$ with $l_i$, we now have  two different thresholds for healthy and converting banks. Thus, at the conversion point there are two potential sources of discontinuity, the shift in the threshold and the shift in the total equity value. We have continuity when the two effects cancel out, i.e., when the difference between the conversion and non-conversion equity values equals the difference of the thresholds. This boils down to $v_i^{C}-v_i^{H} = l_i m_i$, that is $c_i =l_i m_i$, the characterization of the fair case.  Another way to write this condition is as $c_i =v_i^C m_i/(m_i+1)$. At the threshold, owning a fraction $m_i/(m_i+1)$ of the post-conversion equity $v_i^{C}$ is worth the same as receiving $c_i$.

In the remainder of this section, we investigate the comparative statics of the system 
(\ref{eqE1}--\ref{eqE2}). We keep the debt amounts $c_i$ and the weights $w_{ij}$ fixed and restrict attention to the fair case. Moreover, for simplicity, we assume that $m_i=m$ is the same for all banks $i$. Thresholds are thus given by $l_i=c_i/m$. By our previous analysis, there exists a unique equilibrium equity value for all $m\in (0,\infty)$ and all $a\in\mathbb{R}^n$.

We now consider the limit $m\rightarrow \infty$ of the system of equations (\ref{eqE1}--\ref{eqE2}). In this limit, the fraction of the equity value that is handed over to creditors in case of a conversion approaches 100\%. The limit would not have been well-defined for the original system (\ref{eqS1}--\ref{eqS2}). This is the main motivation for switching to the perspective of total equity values.  

In this limit, we have $l_i=0$ for all banks and a combination of asset and equity value vectors $a$ and $v$ together with a partition $(B,C,H)$ form an equilbrium if 
\begin{align}\label{eqEN1a}
 v_i &= a_i + \sum_{j \in C} w_{ij}  v_j + \sum_{j \in H} w_{ij} c_j &\text{for all }i \in B \cup C \\
 v_i &= a_i-c_i + \sum_{j \in C} w_{ij}  v_j + \sum_{j \in H} w_{ij} c_j &\text{for all }i \in H
\label{eqEN2a}
\end{align}
and if $v_i < 0$ for all $i \in B$,  $0 \leq v_i \leq c_i$ for all $i\in C$ and  $v_i > 0$ for all $i\in H$. 

Comparing the system of equations (\ref{eqEN1a}--\ref{eqEN2a}) to the model of bankruptcies in an interbank credit model due to \cite{eisenberg2001systemic}, we find that the two models coincide up to a potentially important terminological difference. What is called a conversion in our model is called a bankruptcy in theirs.\footnote{In \cite{eisenberg2001systemic} and the subsequent literature such as \cite{rogers2013failure}, it is often assumed that $a_i\geq 0$ for all $i$. Then, complete bankruptcies without even partial repayment of debt in the sense of our set $B$ are ruled out. Here, we allow for negative $a_i$ to obtain the complete picture. A bank with negative $a_i$ may well be healthy (or converting) if it receives a sufficient amount of outstanding debt back from its competitors.}

An optimistic reading of this result is as follows: In a world where ordinary debt is replaced by CoCos, bankruptcies are replaced by conversions as long as total asset values are non-negative. Indeed, one might argue that replacing bankruptcies by conversions is more than just a change of terminology. With CoCos, a potentially unpredictable bankruptcy process is replaced by the orderly fulfillment of contractual obligations. Nevertheless, in the limit $m\rightarrow \infty$, complete control of the bank is transferred to new owners which will certainly lead to frictions in practice -- even if the word ``bankruptcy'' is avoided. 

In light of these considerations, it seems worthwhile to consider CoCos with other values of $m$. Intuitively, the character of a conversion changes gradually with $m$. The smaller $m$ is, the less disruptive is a conversion event. In particular, as long as $m\leq 1$, the original owners keep a majority of the stocks so that control of the bank does not change in case of a conversion.\footnote{This reasoning assumes, of course, that the bank's pre-conversion ownership structure is not too fragmented.} 

An additional benefit of choosing a smaller $m$ is that this weakens potential interaction and spillover effects between banks. Formally, when we compare the Eisenberg-Noe equations (\ref{eqEN1a}--\ref{eqEN2a}) to the fair case of (\ref{eqE1}--\ref{eqE2}), we observe two effects. First, the threshold for a bank being healthy is moved upwards from $0$ to $l_i=c_i/m_i$. Similarly, the bound for conversion is moved from $c_i$ to $(1+m_i)l_i=c_i+c_i/m_i$. Yet, second, what this buys us is a decrease in the interconnectedness of the banking system as the weights $w_{ij}$ are decreased to $w_{ij}m_i/(1+m_i)$.

\begin{figure}[htb]
	\begin{center}
	\end{center}
	\begin{center}
		\begin{subfigure}[b]{0.3\textwidth}
			\includegraphics[width=\textwidth]{pic75c8c8.png}
			\caption{$m=1$}\label{fig_EN:a}
		\end{subfigure}
		\begin{subfigure}[b]{0.3\textwidth}
			\includegraphics[width=\textwidth]{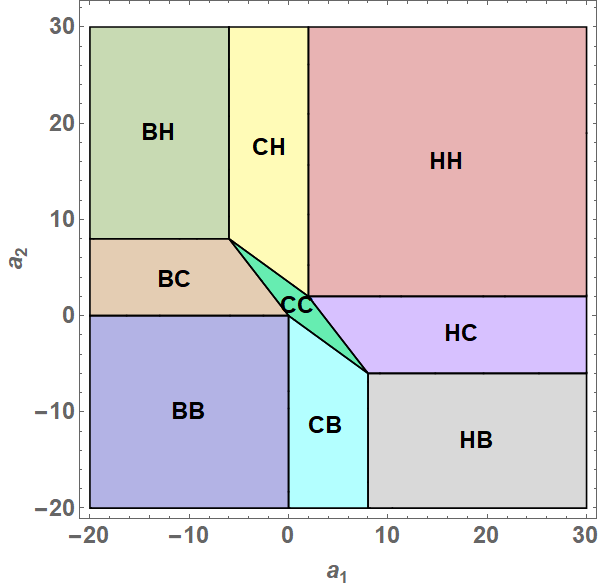}
			\caption{$m\rightarrow\infty$}\label{fig_EN:b}
		\end{subfigure}\\
		\begin{subfigure}[b]{0.3\textwidth}
		\includegraphics[width=\textwidth]{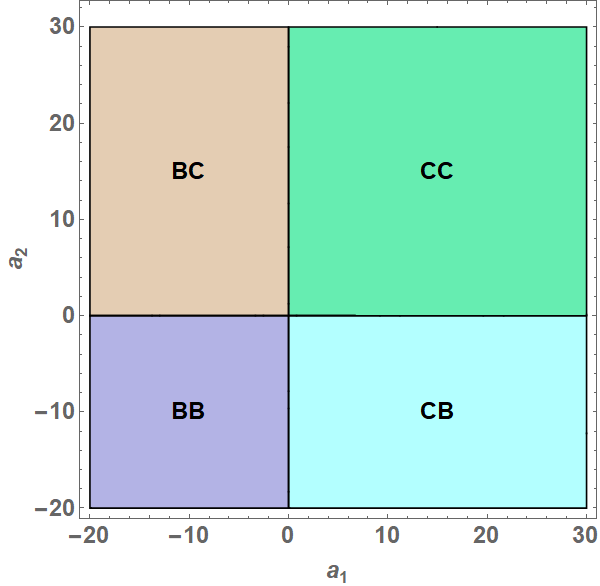}
		\caption{$m\rightarrow0$}\label{fig_EN:c}
		\end{subfigure}
			\begin{subfigure}[b]{0.3\textwidth}
		\includegraphics[width=\textwidth]{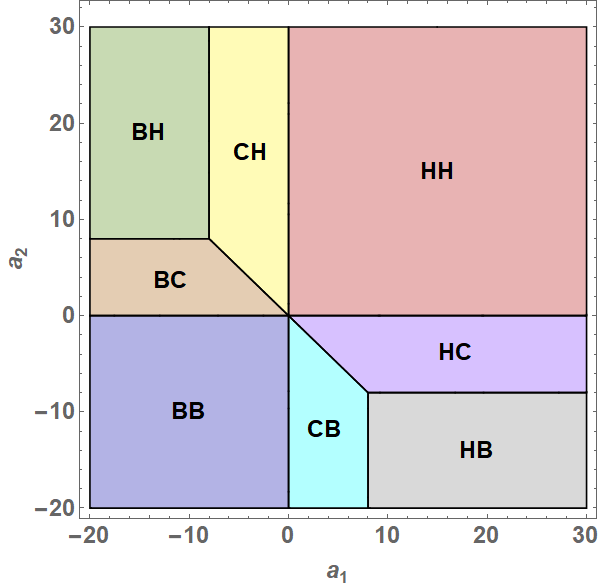}
		\caption{$m\rightarrow \infty$, $w_{12}=w_{21}=1$}\label{fig_EN:d}
	\end{subfigure}
	\end{center}
	\caption{Limiting behavior.}\label{fig_EN}
	\vspace*{0.15cm}
	\parbox{\textwidth}{\footnotesize 
		Equilibria in the fair case for varying $m$ with $c_1=c_2=8$ and, unless otherwise noted, $w_{12}=w_{21}=0.75$.   }
\end{figure}

Visually, when we compare Figures~\ref{fig_EN:a}, \ref{fig_EN:b} and \ref{fig_EN:c} for the case of two banks, we see indeed that regions become more rectangular as $m$ decreases. This means that when the assets of one bank decrease this typically only affects the status of that bank but usually not that of the others.  In the most connected extreme case, $w_{12}=w_{21}=1$ and $m=\infty$, the region where both banks convert degenerates to a decreasing straight line as seen in Figure~\ref{fig_EN:d}. Intuitively, simultaneous conversions imply that the ownership of each bank is transferred fully to the other bank and then on and on in an infinite cycle. This is not possible. Thus the region where both banks are bankrupt touches the region where both are healthy.

To understand the downside of choosing $m$ too small, it is instructive to study the limit $m\rightarrow 0$ depicted in Figure~\ref{fig_EN:c}. In this limit we have $l_i=\infty$ for all banks and a combination of asset and equity value vectors $a$ and $v$ together with a partition $(B,C,H)$ form an equilibrium if 
\begin{align}\label{eqEN1c}
 v_i &= a_i +  \sum_{j \in H} w_{ij} c_j &\text{for all }i \in B \cup C \\
 v_i &= a_i-c_i + \sum_{j \in H} w_{ij} c_j &\text{for all }i \in H
\label{eqEN2c}
\end{align}
and if $v_i < 0$ for $i \in B$ and $0 \leq v_i \leq  \infty$ for $i\in C$. Formally, the condition associated with $i\in H$ becomes $v_i > \infty$ which implies that we must have $H=\emptyset$ in equilibrium. The system thus simplifies to $v_i=a_i$ for all $i$. Bank $i$ converts whenever $a_i \geq 0$. Otherwise, bank $i$ is bankrupt. There are no healthy banks. 

Thus, in the limit $m\rightarrow 0$, replacing debt by CoCos boils down to canceling all debt. No payments are made or received and a bank survives if and only if its assets net of debt $a_i$ are sufficient. In particular, there are no longer any spillovers between banks.  To understand how this situation can go together with our implementation of a ``fair'' conversion threshold, recall that fairness only holds for conversions that occur \textit{at} the threshold. As $m$ approaches 0, the fair conversion threshold goes to infinity and, in the boundary case of a fair conversion, a bank's creditors receive $c_i$ in the form of an infinitesimally small fraction of an infinitely valuable bank. Realistically, as soon as $m$ is sufficiently small, the bank will almost always convert, and it will do so at a value that lies \textit{far} below the fair threshold.

While we have not explicitly modeled the earlier stage at which CoCos are initiated, priced and sold, it seems clear that it will not be possible to raise significant amounts of capital with CoCos whose fair thresholds are too high ($m\rightarrow 0$). Conversely, CoCos whose thresholds are too low will behave more and more similarly to ordinary debt ($m\rightarrow \infty$). In between those extremes, there is a potential scope for CoCos that have at least two advantages over ordinary debt. First, network effects can be expected to be weaker. Second, conversion events will be sufficiently distinct from bankruptcies to be perceived differently by the market. The latter point is in line with results by \cite{chen2017contingent} who study CoCos with accounting-based triggers in a dynamic, single-bank model. They argue that conversion thresholds should be sufficiently high to preclude what they call debt-induced collapse, a regime in which CoCos are effectively reduced to straight debt.

\section{Conclusion}

In this paper, we have studied how trading in contingent convertible debt influences equilibrium stock prices of banks. Our starting point was an ongoing debate about the benefits but also the dangers of such contingent debt. A major concern that has been raised in the previous literature is that equilibrium stock prices may no longer exist or be unique when conversion depends on current values of stock prices. We provide clear-cut conditions that guarantee existence and uniqueness of equilibrium. 

Our results are the first to address the role of contingent convertible debt with stock price triggers in interbank networks. Accordingly, many things remain to be done. First, a natural extension of our setting considers banks that have issued multiple CoCos with different thresholds or different maturities. Second, in order to extend the discussion of Section \ref{sec:disc} into a full welfare analysis,  it would be worthwhile to explicitly model an initial stage where CoCos are issued, priced and sold. This would give more insight into the possibilities for raising capital with different CoCo designs. Third, relatedly, it is possible to study dynamic versions of our model along the lines of \cite{GN} and \cite{PT1} for the single-bank case. Fourth, one could try to characterize the full set of equilibria of the super-fair case.  Finally, it would be interesting to investigate the computation of equilibria similar to the analysis in \cite{schuldenzucker2017complexity} for the case of credit default swaps.\footnote{While we have largely ignored computational aspects in our presentation, our results do have some computational implications. The fixed point construction for the super-fair case implies an explicit algorithm for computing one equilibrium of the super-fair case from equilibria of the fair case. Moreover, the formal resemblance between the fair case and the Eisenberg-Noe model implies that tools can be transferred between these settings. For instance, their ``fictitious default'' algorithm can be translated into a ``fictitious conversion'' algorithm for finding equilibria in the positive quadrant, $a\geq 0$.}

\appendix
\section{Proofs}
\subsection{Proofs of Section \ref{sec:set}}
\begin{myproof}{Lemma}{\ref{lem:ex1}}
The consistency condition in the definition of an equilibrium implies that the stock price vector immediately determines a unique candidate for an equilibrium partition as stated in the lemma. It remains to show that for a given stock price vector $s$ and partition $(B,C,H)$ the linear system (\ref{eqS1}--\ref{eqS2}) possesses a unique solution $a$. This follows from the  discussion following \eqref{eqS3}.
\end{myproof}

\subsection{Proofs of Section \ref{sec:overview}}
\begin{myproof}{Proposition}{\ref{prop:subfair}}
Suppose bank $i$ has set a sub-fair threshold, $l_i m_i<c_i$. To construct a vector of asset values $a\in \mathbb{R}^n$ for which no equilibrium exists, we assume that the asset values of the remaining banks are sufficiently high such that if an equilibrium exists their stock prices must lie above the threshold and we have $j\in H$ for all $j\neq i$. A sufficient condition is that $a_j$ satisfies $a_j -c_j > l_j(1+m_j)$ for all $j \neq i$. Defining $C_i=\sum_{j\neq i} w_{ij} c_j$, the two candidates for the stock price of bank $i$ are then $s_i=\frac{1}{m_i}\left(a_i+C_i \right)$ if $s_i \leq l_i$ and $s_i=a_i-c_i+C_i$ 
if $s_i > l_i$. Plugging the candidate stock prices into the constraints and solving for $a_i$ yields the inequalities $a_i > l_i+c_i-C_i$ and  $a_i \leq l_i(1+m_i)-C_i$. Existence of equilibrium means that at least one of the two inequalities is satisfied for every $a_i \in \mathbb{R}$. Yet, $c_i>l_im_i$ implies $l_i + c_i - C_i > l_i(1+m_i) - C_i$. Thus, there exists an $a_i \in \R$ such that  $l_i + c_i - C_i > a_i > l_i(1+m_i) - C_i$. For this $a_i$, no equilibrium stock price exists.  
\end{myproof}

\subsection{Proofs of Section \ref{sec:fair}}

Our existence proof is based on the Poincar\'e-Miranda Theorem, a classical result from real analysis, see e.g. the Corollary to Proposition 3 in \citet{browder1983}. It is repeated here for the reader's convenience.

\begin{thm}[Poincar\'e-Miranda]\label{tPM}
Define $U_i =\{u \in [-1,1]^n|u_i=1\}$ and $U_{-i} =\{u \in [-1,1]^n|u_i=-1\}$.
Let $F:[-1,1]^n \rightarrow \mathbb{R}^n$ be a continuous function with the property that for all $i\in [n]$ $u\in U_i$ implies $F_i(u)\geq 0$ and $u\in U_{-i}$ implies $F_i(u)\leq 0$. Then there exists $w\in [-1,1]^n$ such that $F_i(w)=0$ for all $i\in[n]$.
\end{thm}

\begin{myproof}{Proposition}{\ref{propIVT}}
The proof relies on Theorem \ref{tPM}. We mainly need to show how to apply this theorem in our context. Notice first that it suffices to prove that there exists $x\in \mathbb{R}^n$ such that $f(x)=0$. The reason is that if $f$ satisfies continuity and properties (i) and (ii) then so does any translation $\widetilde{f}(x)=f(x)-y$ for fixed $y \in\mathbb{R}^n$. Thus, our results apply equally to $f$ and $\widetilde{f}$. Proving existence of $x \in \mathbb{R}^n$ with $\widetilde{f}(x)=0$ implies existence of $x$ with $f(x)=y$.

Next, we have to prove that we can rescale our function to a function on the unit cube which has the boundary conditions required by Theorem \ref{tPM}. By property (ii), there exists a constant $t>0$ such that for all $i\in [n]$ we have $f_i(te)\geq 0$ and $f_i(-te)\leq 0$. Defining $U_i$ and $U_{-i}$ as in  Theorem \ref{tPM}, we conclude from property (i) that $f_i(tu)\geq f_i(te)\geq 0$ for all $u \in U_i$ and $f_i(tu)\leq f_i(te)\leq 0$ for all $u \in U_{-i}$. It follows that the function $F:[-1,1]^n \rightarrow \mathbb{R}^n$ defined by $F(u)=f(t u)$, $u\in [-1,1]^n$ satisfies the requirements of Theorem  \ref{tPM}. In particular, $F$ is continuous with $F_i(u) \geq 0$ for $u \in U_i$ and $F_i(u) \leq 0$ for $u \in U_{-i}$. Thus, by the theorem, there exists a $w \in [-1,1]^n$ with $F(w)=0$. Observing that $f(x)=F(w)=0$ for $x=tw$ concludes the proof.
\end{myproof}

\begin{myproof}{Proposition}{\ref{prop:surj}}
We show that Proposition \ref{propIVT} is applicable with $f \equiv \Phi$. To this end, we need to verify that $\Phi$ is continuous and satisfies properties (i) and (ii). Note first that for any given partition $(B,C,H)$ and associated set of stock prices $\cS_{B,C,H}$ we have
\[
\Phi(s)=L_{B,C}s + b_{H} \;\;\text{ for all } \;\; s\in \cS_{B,C,H}
\]
where the matrix $L_{B,C}$ has positive diagonal elements and non-positive off-diagonal elements. This shows that within each of the sets  $\cS_{B,C,H}$ the function $\Phi$ is continuous and has the monotonicity property required in (i). The global continuity of $\Phi$ which is shown below implies that $\Phi$ also implies (i) globally. We next turn to (ii). We show a slightly stronger claim which covers both cases. Fix some $u\in \{-1,1\}^n$. For $t > \max_{i\in [n]} l_i $, we have that $t u \in \cS_{B,C,H}$ with $B=\{i \in [n]\in u_i=-1\}$, $C=\emptyset$ and $H=\{i \in [n]\in u_i=1\}$. The associated matrix $L_{B,C}$ is a diagonal matrix with positive diagonal entries. Thus, $\Phi_i(t u)$ converges to $+\infty$ for $u_i=1$ and to $-\infty$ for $u_i=-1$.

It remains to verify continuity of $\Phi$. To this end, it suffices to consider the behavior of $\Phi$ at the boundaries between different partition elements. Specifically, we show that if $s\in \overline{\cS_{B_1,C_1,H_1}} \cap \overline{\cS_{B_2,C_2,H_2}}$ for two partitions ${B_1,C_1,H_1}$ and ${B_2,C_2,H_2}$ then the corresponding local definitions of $\Phi$ coincide,
\begin{equation}\label{LeqL}
L_{B_1,C_1}s + b_{H_1}=L_{B_2,C_2}s + b_{H_2}.
\end{equation}
The condition $s\in \overline{\cS_{B_1,C_1,H_1}} \cap \overline{\cS_{B_2,C_2,H_2}}$ means that at least one bank $i$ satisfies $s_i=0$ or $s_i=l_i$ so that the status of that bank is at the boundary between bankruptcy and conversion or between conversion and being healthy. As these conditions are independent between banks, it suffices to show \eqref{LeqL} for pairs of partitions which differ in exactly one bank.\footnote{This is seen easily by inspecting, e.g., the left panel of Figure \ref{figPhi}. To check the condition for the point which lies at the intersection of the (completions of the) sets BH and CC, it suffices to check the condition for the intersections of BH and CH and of CH and CC.} 
We need to consider two cases. In the first one $s_i=0$, $B_2 = B_1 \setminus \{i\}$, $C_2 = C_1 \cup \{i\}$ and $H_2 = H_1$. Let $I_i = e_ie_i^T$ be the matrix with zeros in all entries except for a $1$ in $(i,i)$.
We have $L_{B_2,C_2} = L_{B_1,C_1} - m_iI_i + m_i(I-W)I_i = L_{B_1,C_1} - m_iWI_i$ and $b_{H_2} = b_{H_1}$ and thus
\begin{align*}
(L_{B_2,C_2}s + b_{H_2})&-(L_{B_1,C_1}s + b_{H_1}) \\
&= m_iWI_i s = s_i m_iWe_i = 0.
\end{align*}
In the second case $s_i=l_i$, $B_2 = B_1$, $C_2 = C_1 \setminus \{i\}$ and $H_2 = H_1\cup \{i\}$.
Then $L_{B_2,C_2} = L_{B_1,C_1} - m_i(I-W)I_i$ and $b_{H_2} = b_{H_1} + c_i(I-W)e_i$.
We thus find that in this case
\begin{align*}
(L_{B_2,C_2}s + b_{H_2})&-(L_{B_1,C_1}s + b_{H_1})  \\
&= - m_i(I-W)I_i s + c_i(I-W)e_i\\
&= (c_i - l_i m_i)(I-W)e_i = 0,
\end{align*}
where the final conclusion uses the fairness assumption $l_i=c_i/m_i$.
\end{myproof}

\begin{myproof}{Proposition}{\ref{prop:inj}}
Surjectivity has already been shown in Proposition \ref{prop:surj}. Thus, to prove bijectivity we only to show injectivity. Our proof works in two steps. First, we prove the following \textbf{claim}: Consider two vectors $s$ and $t$ in $\mathbb{R}^n$ such that $s\geq t$ and that the set $I=\{i|s_i>t_i\}$ is non-empty. Then there exists an $i\in I$ such that $\Phi_i(s) > \Phi_i(t)$. Moreover, $\Phi_j(s) \leq \Phi_j(t)$ for all $j \notin I$. Second, we use the claim to complete the proof. To prove the claim, consider first the case $j \notin I$ and thus $s_j=t_j$. By Proposition \ref{prop:surj}, the function $\Phi$ is continuous. Moreover, within each partition element $\mathcal{S}_{B,C,H}$ it is an affine function of the form $L_{B,C} s+b_{H}$ where $L_{B,C}$ has strictly positive diagonal entries and non-positive off-diagonal entries. Thus, $\Phi_j(s)$ is weakly decreasing in $s_i$ for all $i\neq j$ and we have $\Phi_j(s) \leq \Phi_j(t)$ because $s_j=t_j$. To complete the proof of the claim, we show that $e^\top\Phi(s)>e^\top\Phi(t)$, i.e., the sum of the elements of $\Phi(s)$ is strictly larger than the sum of the elements of $\Phi(t)$. To see this, it suffices to recall the continuity of $\Phi$ and the local definitions of $\Phi$ and to note that the vectors $e^\top L_{B,C}$ are non-negative with some strictly positive entries because $e^\top (I-W) \geq 0$. Thus, there must exist an $i$ with  $\Phi_i(s) >\Phi_j(t)$ and, by the other part of the claim, this $i$ must lie in the set $I$.

It remains to show that the \textbf{claim} implies injectivity. Consider two vectors $s$ and $u$ in $\mathbb{R}^n$ with $s\neq u$. We need to show that $\Phi(s)\neq \Phi(u)$. To this end, define $t=\min(s,u)\in \mathbb{R}^n$ where the minimum is taken entrywise. Define $I_s=\{i| s_i >t_i\}$ and $I_u=\{i| u_i >t_i\}.$ Note that $I_s$ and $I_u$ are disjoint and that at least one of them is non-empty. Without loss of generality, assume $I_s \neq \emptyset$. Applying the claim to $s$ and $t$ shows that there exists an $i^*\in I_s$ such that $\Phi_{i^*}(s) > \Phi_{i^*}(t)$. Moreover, we know that  $i^*\notin I_u$, i.e., $u_{i^*}=t_{i^*}$. Thus, applying the claim to $u$ and $t$ implies $\Phi_{i^*}(u) \leq \Phi_{i^*}(t)$ and thus $\Phi_{i^*}(u)\neq \Phi_{i^*}(s)$. This completes the proof.
\end{myproof}

\subsection{Proofs of Section \ref{sec:superfair}}

We begin this section with a technical lemma. Lemma \ref{lem:fixpoint} is proved at the end. The remaining results are proved in the order in which they are stated in the main text. 

\begin{lem}\label{lem:M-matrix} If $I-W$ is invertible, the matrix $(I-W)^{-1}$ has only non-negative entries. Moreover, for any $B,C \subseteq [n]$ the matrix $L_{B,C}^{-1}(I-W)$ has non-negative entries on the diagonal and non-positive entries otherwise.
\end{lem}
\begin{myproof}{Lemma}{\ref{lem:M-matrix}}
For any $\alpha \in(0,1)$, the matrix $I-\alpha W$ is strictly column-diagonally dominant with positive diagonal entries and non-positive off-diagonal entries. It is thus an $M$-matrix which implies that its inverse $(I-\alpha W)^{-1}$ has only non-negative entries, see Chapter 2.5 of \cite{roger1994topics}. As a limit $\alpha \rightarrow 1$ of non-negative matrices, $(I-W)^{-1}$ is then also non-negative. For the second claim, notice that $L_{B,C}^{-1}(I- W)$ is the limit $\alpha \rightarrow 1$ of the matrices
\[
M_\alpha = \left(I + \Diag(m_B)  + (I- \alpha W) \Diag(m_C) \right)^{-1}(I-\alpha W)
\]
so it suffices to show that $M_{\alpha}$ has the required sign-pattern. To this end, we write
\[
M_\alpha = \left( (I-\alpha W)^{-1} (I + \Diag(m_B))  + \Diag(m_C) \right)^{-1}.
\]
As argued before, for $\alpha \in (0,1)$, $(I-\alpha W)^{-1}$ is the inverse of an $M$-matrix. By Theorem 1 and 3 of \citet{johnson1982inverse}, the  family of inverse $M$-matrices is closed under multiplication by diagonal matrices with positive diagonal and under addition by diagonal matrices with non-negative diagonal. Thus, $M_\alpha^{-1}$
is an inverse $M$-matrix. Consequently $M_\alpha$ itself is an $M$-matrix which means that it has the required sign pattern.
 \end{myproof}

\begin{myproof}{Lemma}{\ref{lem:preserveSomeHealth}}
Consider the segment $a(t) = a + t(I-W)d_X$ where $t\in [0,1]$. As $\Phi$ is a continuous piece-wise linear bijection we have that $s(t) =\Phi^{-1}(a(t))$ is a piece-wise linear path in $\R^n$. Moreover, there are $0=t_0<t_1<\cdots<t_N=1$ such that the partition $(B(t),C(t),H(t))$ of $[n]$ corresponding to the equilibrium $(s(t),a(t))$  is constant in the interval $(t_{k-1},t_k)$ for each $k= 1,\dots,N$. We denote this partition $(B_k,C_k,H_k)$.

Let $1 \le k \le N$. For any $t \in (t_{k-1},t_k)$, there are $s_k,u_k \in \R^n$ such that $s(t) = s_k + t u_k$. The identity $a(t) = \Phi(s(t))$ then becomes $a + t(I-W)d_X = L_{B_k,C_k}(s_k + t u_k) + (I-W)c_{H_k}$. This implies $(I-W)d_X = L_{B_k,C_k} u_k$, and therefore $u_k = L_{B_k,C_k}^{-1}(I-W)d_X$. 
From Lemma~\ref{lem:M-matrix},  the row vector $e_i^TL_{B_k,C_k}^{-1}(I-W)$ has non-positive  entries, except for a non-negative entry in position $i$. On the other hand, if $i\notin X$, $(d_X)_i = 0$. Thus $(u_k)_i = e_i^T u_k = e_i^TL_{B_k,C_k}^{-1}(I-W)d_X \le 0$  for all $i \notin X$. This implies that $H_{k+1} \subseteq H_k \cup X$ and, using $H_1 = H(a)$, we obtain by induction that $H(a(t)) \subseteq H(a) \cup X$ for all $t \in [0,1]$. By taking $t=1$, the claim of the lemma follows.
\end{myproof}

\begin{myproof}{Lemma}{\ref{lem:presSomeHealth3}}
First notice that without loss of generality we can assume $d_j > 0$ for all $j \in X$. Otherwise we can replace $X$ by $\{j \in X: d_j > 0\}$.   We prove the statement by induction over the size of $X$.
To this end, we will construct $f \in \R^n$ such that $0 \le f \le d_X$,  $f_j = d_j$ for some $j \in X$ and $X\subseteq H(a + (I-W)f)$.
Given such an $f$, we can define  $X' = \{j\in X: d_j > f_j\} \subsetneq X$. If $X' = \emptyset$ then we must have $f = d_X$ and therefore $X \subseteq H(a+(I-W)d_X)$. Otherwise, applying the induction hypothesis to $a' = a + (I-W)f$, $d' = d - f$, and $X'$, we obtain $\emptyset \neq X' \cap H(a' + (I-W)d'_{X'})$. Since $d'_{X'} = (d-f)_{X'} = d_X - f$, we can thus write $a'+ (I-W)d'_{X'}  = a+(I-W)f + (I-W)(d_X - f) = a + (I-W)d_X.$ It follows that $X \cup H(a + (I-W)d_X) \supseteq X' \cap H(a'+(I-W)d'_{X'}) \neq \emptyset$. Notice that this argument also covers the induction basis where $X$ is a singleton so that $X' = \emptyset$.

To complete the proof we need to show that such an $f$ exists. Let $u = (I-W)^{-1}d_X$. We know that $u \ge 0$, as $(I-W)^{-1}$ is a non-negative matrix by Lemma \ref{lem:M-matrix}. Moreover, for any $j \in X$ we have $0 < d_j = e_j^Td_X = e_j^T(I-W)u$. Yet $e_j^T(I-W)$ is non-positive, except for the $j$-entry. Thus $u_j >0$.  Let $t = \min_{j\in X} \tfrac {d_j}{u_j}$ and let $f =  tu_X$. Then, we have $0 \le f \le d_X$ and $f_j = d_j$ for some $j \in X$.  It remains to show that $X \subseteq H(a + (I-W)f)$. First, we argue that  $H(a+t(I-W) u) = H(a)$. To prove this let $(B,C,H)$ be a partition  of $[n]$ such that $(a,\Phi^{-1}(a))$ is a $(B,C,H)$-equilibrium. As $X \subseteq H(a) = H$, we have $L_{B,C}(\Phi^{-1}(a) + td_X) + c_H = a + td_X$ so that  $(a+ td_X,\Phi^{-1}(a) + td_X)$ is also a $(B,C,H)$-equilibrium. This implies $H(a+ t(I-W)u ) = H(a+td_X) = H = H(a)$. From Lemma~\ref{lem:preserveSomeHealth}, it then follows that $X \subseteq H(a) = H(a+t(I-W)u) = H(a+t(I-W)u_X + t(I-W)u_{[n]\setminus X}) \subseteq H(a+(I-W)f) \cup ([n]\setminus X)$. Thus $X \subseteq H(a+(I-W)f)$.
   
\end{myproof}

\begin{myproof}{Corollary}{\ref{cor:antimon1}}
\begin{enumerate}[(i)]
  \item Assume $X \subseteq Y \subseteq  h(X)$. Then $h(Y) = H(\hat a + (I-W)d_X + (I-W)d_{Y\setminus X}) \subseteq h(X) \cup (Y\setminus X) =  h(X)$, from Lemma~\ref{lem:preserveSomeHealth}.
  \item Assume $ h(X) \subseteq Y \subseteq X$. Then $h(X) = H(\hat a + (I-W)d_X) \subseteq  H(\hat a + (I-W)d_X - (I-W)d_{X\setminus Y}) \cup (X\setminus Y) = h(Y) \cup (X\setminus Y)$, from Lemma~\ref{lem:preserveSomeHealth}. But, $h(X) \cap (X\setminus Y) = \emptyset$ and thus $h(X) \subseteq h(Y)$.
 \item Assume $ X \subseteq h^2(X) \varsubsetneq h(X)$. Taking $Y= h^2(X)$ in (\ref{loc:antimon1}) we obtain $h^3(X)\subseteq h(X)$. Applying (\ref{loc:antimon2}) to $h^2(X) \subseteq h^2(X) \subseteq h(X)$,  we obtain $h^2(X) \subseteq h^3(X)$. Now we need to show $h^3(X)\neq h(X)$. For the sake of contradiction, assume  $h^3(X) = h(X)$. Then
     $\emptyset \neq h(X) \setminus h^2(X) \subseteq h^3(X) = H(\hat a + (I-W)d_{h^2(X)})$. From Lemma~\ref{lem:presSomeHealth3}, we have then that
     $\emptyset \neq (h(X) \setminus h^2(X)) \cap H(\hat a + (I-W)d_{h^2(X)} + (I-W)d_{h(X) \setminus h^2(X)})= (h(X) \setminus h^2(X)) \cap h^2(X) = \emptyset$.
 \item Assume $ h(X) \varsubsetneq h^2(X)\subseteq X$. Taking $Y= h^2(X)$ in (\ref{loc:antimon2}) we obtain $h(X)\subseteq h^3(X)$. Applying (\ref{loc:antimon1}) to $h(X) \subseteq h^2(X) \subseteq h^2(X)$  we obtain $h^3(X) \subseteq h^2(X)$. Now we need to show $h^3(X)\neq h(X)$. Notice that  $\emptyset \neq h^2(X) \setminus h(X) \subseteq h^2(X) = H(\hat a + (I-W)d_{h(X)})$. From Lemma~\ref{lem:presSomeHealth3}, we have then that
     $\emptyset \neq (h^2(X) \setminus h(X)) \cap H(\hat a + (I-W)d_{h(X)} + d_{h^2(X) \setminus h(X)})= (h^2(X) \setminus h(X)) \cap h^3(X) = (h^2(X) \cap h^3(X)) \setminus h(X) = h^3(X) \setminus h(X)$.
\end{enumerate}
\end{myproof}

\begin{myproof}{Lemma}{\ref{lem:fixpoint}}
Let $X_m = h^m(H(\hat{a}))$, for all $m \ge 0$ where $\hat{a}\in \mathbb{R}^n$ is arbitrary and fixed and $h$ is defined in the main text. We claim that for some $m$ we have $X_{m+1} = X_m$. $X_m$ is then the desired fixed point as $X_m = H(\hat a + (I-W)d_{X_m})$ by the definition of $h$.  First notice that $X_1 \subseteq X_0$ as  $X_1 = H(\hat a + (I-W)d_{X_0}) \subseteq H(\hat a) \cup X_0 = X_0$ by Lemma \ref{lem:preserveSomeHealth}. Similarly, it follows that $X_1 \subseteq X_2 \subseteq X_0$: As $h(X_0) = X_1 \subseteq X_0$, from Corollary~\ref{cor:antimon1}.(\ref{loc:antimon2}), we have $X_1 \subseteq X_2$. Moreover,  $X_2 = H(\hat{a} + (I-W)d_{X_1}) \subseteq H(\hat{a}) \cup X_1 = X_0$.

Now inductively by applying interactively items~(\ref{loc:antimon3}) and~(\ref{loc:antimon4}) from Corollary~\ref{cor:antimon1}, we obtain that $X_m \subseteq X_{m+2} \subseteq X_{m+1}$ and, either $X_m = X_{m+1}$ or $|X_{m+1} \setminus X_{m+2}| < |X_{m+1} \setminus X_m|$, for all odd $m$. And for all even $m > 0$,  $X_{m+1} \subseteq X_{m+2} \subseteq X_{m}$ and,  either $X_m = X_{m+1}$ or $|X_{m+2} \setminus X_{m+1}| < |X_{m} \setminus X_{m+1}|$. This implies that $X_{m+1} = X_m$ for some $m$ as $|X_{0} \setminus X_{1}|\leq n$.
\end{myproof}

 \end{document}